\documentclass[12pt,preprint]{aastex}
\begin{document}

\title {The Brightest Stars in M32: Comparing Predictions from Spectra with the Resolved Stellar Content \altaffilmark{1} \altaffilmark{2}}

\author{T. J. Davidge \altaffilmark{3}}

\affil{Herzberg Institute of Astrophysics,
\\National Research Council of Canada, 5071 West Saanich Road,
\\Victoria, B.C. Canada V9E 2E7\\ {\it email: tim.davidge@nrc.ca}}

\author{Joseph B. Jensen}

\affil{Gemini Observatory, 670 North A'ohoku Place,
\\Hilo, HI 96720-2700\\ {\it email: jjensen@gemini.edu}}

\altaffiltext{1}{Based on observations obtained at the
Gemini Observatory, which is operated by the Association of Universities
for Research in Astronomy, Inc., under a co-operative agreement with the
NSF on behalf of the Gemini partnership: the National Science Foundation
(United States), the Particle Physics and Astronomy Research Council
(United Kingdom), the National Research Council of Canada (Canada),
CONICYT (Chile), the Australian Research Council (Australia), CNPq (Brazil),
and CONICET (Argentina).}

\altaffiltext{2}{This publication makes use of data products
from the Two Micron All Sky Survey, which is a joint project of the University of
Massachusetts and the Infrared Processing and Analysis Center/California 
Institute of Technology, funded by the National Aeronautics and Space Administration 
and the National Science Foundation.}

\altaffiltext{3}{Visiting Astronomer, Canada-France-Hawaii
Telescope, which is operated by the National Research Council of Canada,
the Centre National de la Recherche Scientifique, and the University of
Hawaii.}

\begin{abstract}

	Broad- and narrow-band images covering the $1 - 4\mu$m wavelength interval 
are used to investigate the properties of the brightest AGB stars in the Local Group 
galaxy M32. Data obtained with the NIRI imager on the Gemini North telescope
indicate that the brightest AGB stars near the center of M32 have peak 
M$_{L'}$ brightnesses and $K-L'$ colors that are similar to those of luminous 
AGB stars in the Galactic disk. Data obtained with the CFHTIR imager on the 
Canada-France-Hawaii Telescope indicate that the density of bright AGB stars per unit 
visible and near-infrared surface brightness is constant out to projected major axis 
distances of 1 kpc, suggesting that the AGB stars and their progenitors are 
smoothly mixed throughout the main body of the galaxy. In 
addition, the $J-K$ color distribution of bright AGB stars throughout much of the galaxy 
is consistent with that of a single population of AGB stars, the majority of which are 
long period variables, having a common metallicity and age. 
Thus, these data do not support spectroscopic studies that find an age gradient in M32. 
The AGB contributes $70^{+30\%}_{-20\%}$ of the 
integrated light in the region surveyed. This is consistent with previous estimates made 
from the integrated near-infrared spectrum, and is suggestive of an age $\sim 2$ 
Gyr. The stellar content of M32 is compared with that 
of the M31 bulge at a projected minor axis distance of 1.4 kpc. 
While the peak $K-$band brightnesses of AGB stars in the two systems 
agree to within a few tenths of a magnitude, M32 contains 
more bright AGB stars per unit integrated brightness than the outer bulge of M31. 
This is suggestive of a difference in mean age, and it is concluded that the star forming 
histories of M32 and the bulge of M31 have differed over a significant fraction of 
their lifetimes, which is consistent with spectroscopic studies of 
these systems. The well-mixed AGB content of M32 is consistent with the galaxy 
having been tidally stirred, presumably by interactions with M31.

\end{abstract}


\section{INTRODUCTION}

	The Local Group galaxy M32 is the nearest system with structural characteristics 
that are reminiscent of classical elliptical galaxies (e.g. Kormendy 1985; Graham 2002). 
These similarities notwithstanding, the evolution of M32 has almost certainly 
been influenced by -- if not dominated by -- the gravitational field 
of M31. A number of studies have suggested that the compact nature of M32 is the 
consequence of tidal pruning (e. g. Faber 1973; Burkert 1994; Bekki et al. 2001), and 
the orbit of M32 is such that there have likely been interactions with M31 
and some of its companions (Cepa \& Beckman 1988).  

	Graham (2002) and Choi, Guhathakurta, \& Johnston (2002) find evidence of a 
residual disk around M32, suggesting that the present-day galaxy may be the remnant 
of what was once a larger disk-dominated galaxy; if this interpretation is correct then the 
majority of stars that belonged to the original disk may have been 
donated to M31. In fact, there are signatures of interactions between M31 
and at least some of its companions. While M31 likely contains a 
metal-poor halo with properties that are similar to the Milky-Way halo (Chapman 
et al. 2006), the extraplanar environment is dominated by 
tidal streams (e.g. Ferguson et al. 2002; Zucker et al. 2004). 
The orbit of the dominant stream may come close to the center of M31, where 
the stream can be disrupted and the stars dispersed into the inner halo (Ibata et al. 2004).
Tidal stirring may explain the uniform metallicity distribution 
seen throughout the disk and extraplanar regions of M31 (Bellazzini et al. 2003). 

	Ferguson et al. (2005) investigated deep CMDs of tidal substructures 
in the outer regions of M31, and these data reveal two 
common characteristics. First, the red giant branch (RGB) sequences in the substructures 
have similar colors, which are suggestive of [Fe/H] between --0.4 and --0.7. 
Second, the substructures contain a population of blue main 
sequence stars, with a MSTO that is indicative of an age $\geq 2.5$ Gyr. 
The age and metallicities of the structures in the outer regions of M31 are 
thus reminiscent of those inferred for M32 (see below), although the mean velocity and 
velocity dispersion of stars in the dominant stream are difficult to reconcile with 
the stream originating in M32 (Ibata et al. 2004). 

	Other components of M31 and M32 may also have been affected by interactions 
between these galaxies. Beasley et al. (2005) find 
that a subset of M31 globular clusters have ages and chemical 
properties that are suggestive of forming as part of M32. Moreover, the orbital properties 
of M32 are consistent with a link to an HI cloud near M31 (Cepa \& Beckman 1988), 
raising the possibility that at least some of the ISM has been stripped from this galaxy.
Finally, warping in the HI disk of M31 can also 
be explained as the result of interactions with M32 (Byrd 1978). 

	While the evolutionary status of M32 in the context of more massive spheroidal 
systems is not clear, it is still recognized as an important laboratory for 
stellar population studies. Because of its high central surface brightness it is 
possible to obtain high quality spectra that span a wide range 
of wavelengths and can serve as the basis for detailed analysis of the 
integrated light. M32 is also close enough that it can be 
resolved into stars, and there is an absence of interstellar dust, so that differential 
extinction is not an issue when investigating the resolved stellar content. Thus, 
M32 is one of only a handful of galaxies where there is a possibility of comparing 
predictions from the spectroscopic analysis of integrated light with the resolved 
stellar content. 

	Most studies of the integrated spectrum of M32 have found signatures of an 
intermediate age population (O'Connell 1980; Rose 1985; Davidge 1990; Bica, Alloin, \& 
Schmidt 1990; del Burgo et al. 2001; Worthey 2004; Rose et al. 2005). 
The integrated visible and near-infrared colors of M32 show little or no variation 
with radius (e.g. Peletier 1993). While it was initially thought that the far-UV color 
profile of M32 may differ from that of other spheroids (O'Connell et al. 1992; Ohl et al. 
1998), this result was likely a consequence of contamination from the outer disk of 
M31, and de Paz et al. (2005) conclude that the radial UV properties of M32 are similar to 
those of classical ellipticals. There are modest radial trends in the strengths of 
some absorption lines (Davidge, de Robertis, \& Yee 1990; Davidge 1991; Hardy et al. 1994; 
Fisher, Franx, \& Illingworth 1995; Worthey 2004; Rose et al. 2005). As discussed by 
Davidge (1991), Worthey (2004), and Rose et al. (2005), the modest gradients in 
the spectroscopic and photometric data may not mean that the stellar content of M32 is 
uniformly mixed; rather, the radial properties of M32 can be 
explained as the result of gradients in both age and metallicity, in the sense 
of an older mean age and lower mean metallicity towards increasing radii. 

	A spectroscopically and photometrically 
distinct nucleus might be expected in M32 given the presence 
of a supermassive black hole (Tonry 1987; Dressler \& Richstone 
1988). However, there is only modest -- if any -- nuclear line emission 
(e.g. Rose et al. 2005), and the spectral-energy distribution (SED) of the 
nucleus does not differ markedly from that of its surroundings. WFPC2 images show 
no evidence for central sub-structures due to dust or a disk, and there is 
no evidence for a central blue core (e.g. Lauer et al. 1998). 
van der Marel et al. (1998) and Ho, Terashima, \& 
Ulvestad (2003) note that the low level of nuclear emission is surpising 
given the high central stellar density. The absence 
of emission might be related to the low HI and molecular gas content, as measured 
by Sage, Welch, \& Mitchell (1998) and Welch \& Sage (2000). 

	The bright resolved stellar content of M32 has been the subject of 
a number of studies. Davidge et al. (2000) and Davidge \& Rigaut (2004) discuss 
the properties of asymptotic giant branch (AGB) stars between 2 and 13 arcsec of the 
nucleus. Davidge et al. (2000) find (1) no obvious trend in the peak brightness 
or $H-K$ color of bright AGB stars between 2 and 13 arcsec of the nucleus, 
and (2) that the number of bright AGB stars scales with 
$r-$band surface brightness, suggesting that the brightest AGB stars belong to a 
population that is uniformly mixed with the main body of stars. The fraction of 
AGB stars that are long period variable (LPVs) does not change with radius near the 
nucleus (Davidge \& Rigaut 2004). 

	The properties of RGB and AGB stars outside of the central regions of M32 have been 
investigated at visible/red (Freedman 1989; Davidge \& Jones 1992, \& 
Grillmair et al. 1996) and near-infrared (Freedman 1992; 
Elston \& Silva 1992; Davidge 2000) wavelengths. 
Grillmair et al (1996) sampled stars as faint as the horizontal branch, and did not 
find evidence for an extended AGB; rather, they argue that their 
data are consistent with a moderately old stellar population. However, Freedman (1989),
Elston \& Silva (1992), and Davidge (2000) find bright AGB stars at 
near-infrared wavelengths, although it is not clear if such stars come from an 
intermediate-age or an old metal-rich population (e.g. Davidge 2001). 
The seeming discrepancy between the AGB contents found by Grillmair et al. (1996) 
and the near-infrared studies can be understood if 
the photometric properties of the brightest cool stars 
at wavelengths shortward of $1\mu$m are affected by line blanketing 
and circumstellar extinction; in this case, 
a better measure of the luminosity of AGB stars is obtained in the infrared. 

	The deepest photometric study of M32 to date was conducted by Worthey et al. 
(2004), who used WFPC2 images to search for the main sequence turn-off (MSTO) in the outer 
regions of the galaxy. The MSTO was not detected, and it was concluded that 
the youngest population in this part of the galaxy has an age of 1 Gyr or older. 
The detection of RR Lyrae stars in the outer regions of M32 by Alonso-Garcia, Mateo, 
\& Worthey (2004) indicate that an old stellar population is present. 

	The bright stellar content in the outer regions of M32 is 
similar to that in the outer regions of other nearby, but more massive, 
elliptical galaxies. The metallicity distribution 
functions of the outer regions of M32 and NGC 5128 are similar (Harris \& Harris 2000), 
and the brightest AGB stars in these -- and other -- galaxies have comparable peak M$_K$ 
(Davidge 2002). After normalizing to a common surface brightness, the $K-$band LF of AGB 
stars in the outer regions of M32 matches the LFs of the outer regions of Maffei 1 and 
NGC 5128 (Davidge 2002). Clearly, it is of interest to compare the bright stellar 
content of M32 with the bulge of M31. Davidge (2001) compared AGB 
stars in the {\it outer} regions of M32 and the {\it inner} 
bulge of M31. While differences in stellar content were found, the 
central regions of many bulges harbour star clusters that may not be representative 
of the main body of the bulge. 

	Existing studies of the resolved stellar content of M32 have 
typically been restricted to relatively small areas, and this can frustrate efforts to 
identify a sample of AGB stars that is large enough to permit the peak AGB 
brightness to be measured, especially in low density regions at large radii. 
Moreover, data that cover only small areas are of limited use for assessing if there 
are radial changes in stellar content. Contamination from 
stars in the outer disk of M31 is an additional source of potential grief. 
An especially insidious aspect of this contamination is that 
the mean metallicity of stars in the outer disk of M31 is [M/H] 
$\sim -0.6$ (Bellazzini et al. 2003), which is not greatly different from that of the 
main body of stars in M32. As a result, the photometric properties of stars in 
M32 and the outer disk of M31 overlap, complicating efforts to identify stars in 
the portion of M32 that is closest to M31.

	In the current study, two datasets are used to investigate 
the photometric properties of the brightest stars in M32. 
In one dataset, $L'$ images obtained with the NIRI imager on Gemini North are 
combined with the near-diffraction limited $K-$band images used by Davidge et al. 
(2000) to investigate the most highly evolved AGB stars near the center of M32. Data 
longward of $2.5\mu$m are of interest for such an investigation as they can be used 
to identify AGB stars that are embedded in thick circumstellar 
envelopes. These stars are expected to have massive progenitors, and hence probe 
the young end of the `intermediate age' range.

	The other dataset consists of broad- and narrow-band 
images obtained with the CFHTIR on the Canada-France-Hawaii Telescope (CFHT), 
and is used to probe the bright stellar content of M32 over a range of radii. 
In addition to sampling the brightest AGB stars at a wavelength where line blanketing 
is not a major issue, broad-band near-infrared colours can also be used to isolate 
C stars (e.g. Davidge 2003; 2005, Demers, Dallaire, 
\& Battinelli 2002), which might be present if the outer regions 
of M32 contain an intermediate-age population. Finally, images of 
the bulge of M31, obtained during the same observing run with the same instrument, 
are also used to make a direct comparison with the stellar content of M32. 

	The paper is structured as follows. The observations and the 
procedures used to reduce the data are discussed in \S 2. The color-magnitude 
diagrams (CMDs) and luminosity functions (LFs) obtained 
from the NIRI and CFHTIR data are presented and discussed in \S\S 
3 and 4, while the stellar content of M32 and the bulge of M31 are 
compared in \S 5. A discussion and summary of the results follows in \S 6.

\section{OBSERVATIONS \& REDUCTIONS}

\subsection{NIRI Data}

	Deep $L'$ images of M32 were recorded with NIRI (Hodapp et al. 
2003) on Gemini North during the nights of August 16 and September 18, 2003 
as part of program GN-2003B-Q-43. The f/14 camera was used for these 
observations, and a 512${\times}$512 subarray was read out; therefore, 
each exposure covers 25.6 arcsec on a side with 0.05 arcsec pixel$^{-1}$ 
sampling. Hundreds of short (0.076 and 0.151 s) exposures were co-added to produce 
an image with an integration time of 30.2 s at each dither position. A sky field 
was observed immediately after each M32 dither position was completed, and the sky 
and M32 observations have the same exposure times. The dither sequence was repeated 
to build up signal, and the total on-source integration time is 1557.4 s.

	The sky frames that bracket the observations of each M32 dither position were 
averaged and subtracted from those M32 observations. The sky-subtracted data were then 
divided by a flat field frame, which was constructed by normalizing the mean of all 
sky images. Flat-fielding is often ignored when reducing imaging data recorded in the 
thermal infrared because the background noise usually far exceeds flat field variations. 
However, this step was judged to be necessary for these data because the background noise 
is comparable to the flat field variations. The flat-fielded data were 
spatially aligned and then summed. A final image was produced 
by cropping the summed image to the area that is common to all 
dither positions. The image quality in the final $L'$ image is 0.35 arcsec FWHM.

\subsection{CFHTIR Data}

	Near-infrared images of M32 were recorded on the night of November 22, 2002 
with the CFHTIR camera, which was mounted at the Cassegrain focus of the 
3.6 meter CFHT. The detector in CFHTIR is a $1024 \times 1024$ 
HgCdTe array. With a pixel scale of 0.211 arcsec pixel$^{-1}$, then a 
$3.6 \times 3.6$ arcmin$^2$ field is covered with each exposure.

	A field in the southern half of M32, centered at 00:42:42.3 Right Ascension 
and $+$40:50:00 Declination (J2000), was observed. The southern portion of M32 
was selected for this study to minimze contamination from stars in the disk of M31, which 
complicate efforts to study the northern half of the galaxy. A field located near the 
minor axis of M31, and centered at 00:42:30.0 Right 
Ascension and $+$41:19:25 Declination (J2000), was also observed to allow the 
properties of AGB stars in the outer bulge of M31 to be compared with those in M32. 
The locations of both CFHTIR fields are marked in Figure 1; 
the south east corner of the CFHTIR M31 field samples a portion of the bulge that is 
close to the F174 field discussed by Stephens et al. (2003). 

	The CFHTIR fields were observed through $J, H,$ and $K'$ broad-band filters, 
as well as CO and $K-$continuum narrow-band filters. The CO filter samples the 
$(2-0)$ bandhead, with $\lambda_{cen} = 2.30\mu$m and $\Delta \lambda = 0.02\mu$m, 
while the continuum filter has $\lambda_{cen} = 2.26\mu$m and $\Delta \lambda = 0.06\mu$m. 
For the observations taken through the broad-band filters, three 30 s exposures 
were recorded at four dither positions; hence, the total 
exposure time is 360 s filter$^{-1}$. A longer exposure time was 
employed when observing through the narrow-band filters; six 30 s exposures were
recorded per dither position through the continuum filter, and nine 30 s exposures 
were recorded per position through the CO filter.

	The data were processed using a pipeline for near-infrared imaging that 
included (1) dark subtraction, (2) the division by 
dome flats, (3) the subtraction of calibration frames that removed thermal emission 
signatures and interference fringes, and (4) the subtraction of the DC sky level.
The resulting images for each field$+$filter combination were spatially 
registered and then median combined. Final images were produced 
by trimming the stacked images to the area common to all exposures. 
Stars in the final images have FWHM $\sim 0.8$ arcsec, with 
modest filter-to-filter scatter.

\section{LUMINOUS AGB STARS NEAR THE CENTER OF M32}

	The analysis of integrated spectra at visible wavelengths suggests that the 
youngest and most metal-rich stars in M32 are more centrally concentrated than 
older metal-poor stars (e.g. Worthey 2004; Rose et al. 2005). The radial 
changes in stellar content predicted from the spectra are such that measureable 
changes in the properties of resolved stars might be expected near the center of M32. 
For example, Worthey (2004) finds that the mean age is $\sim 4$ Gyr and [M/H] $= +0.05$ 
near the nucleus, whereas at 44 arcsec ($\sim 170$ parsecs) the mean age is $8 - 10$ Gyr 
with [M/H] $= -0.25$. Rose et al. (2005) finds a mean age $3 - 4$ Gyr near the 
nucleus, and $6 - 7$ Gyr at 30 arcsec ($\sim 115$ parsecs) radius. Age 
gradients of the size inferred from the spectra could change the peak AGB brightness 
and/or the density of bright AGB stars over angular scales of a few tens of arcsec.

	Efforts to study the youngest AGB stars in M32 may be complicated by circumstellar 
envelopes around these objects. This is a concern because the youngest, most metal-rich 
AGB stars will have the most massive progenitors, and these objects will likely 
experience the greatest rates of mass loss during their final stages of evolution. 
These objects are then more likely to be located in 
circumstellar cocoons than older, more metal-poor AGB stars. Because of the 
wavelength dependence of dust extinction, and the possible 
heating of dust by the central star, objects in such circumstellar shells may only 
stand out clearly against the main body of stars at wavelengths longward of $2.5\mu$m. 
Given that stars of this nature may be difficult to detect in the near-infrared, 
it was decided to investigate the brightnesses of AGB stars near the center of M32 in 
$L'$, which is a wavelength regime where sources with characteristic temperatures of a 
few hundred K will be brightest.

\subsection{Identifying Unblended Stars and the $(L', K-L')$ CMD}

	The angular resolution of the $L'$ data is 0.35 arcsec FWHM. This is poorer than 
in previous studies of the center of M32 where individual AGB stars have been resolved, 
and crowding may affect the ability to measure the 
brightnesses and colors of even the brightest stars. Following the 
procedure described by Davidge, Jensen, \& Olsen (2006), a sample of objects that is 
likely not significantly affected by blending was identified by investigating the 
effects of degrading the angular resolution on measured brightness. 
This requires an infrared image that has a higher angular resolution than the $L'$ 
data, and the Hokupa'a (Graves et al. 1998) $+$ QUIRC 
$K-$band image, which has FWHM $= 0.12$ arcsec and was 
used by Davidge et al. (2000) to study the AGB content near the center of M32, 
was employed for this purpose. 

	The Hokupa'a $K-$band image was rotated and re-sampled to match the orientation 
and pixel sampling of the NIRI $L'$ image. The result was convolved
with a gaussian so that the angular resolution matched that of the $L'$ image. 
The brightnesses of sources in both the original and smoothed images were measured using 
the PSF-fitting program ALLSTAR (Stetson \& Harris 
1988). The difference in brightness between the smoothed and 
unsmoothed images, D$_K$, is a measure of the impact of blending on a particular 
star; stars where smoothing affects the measured brightness will have D$_K >> 0$, whereas 
sources that are not greatly affected by blending at the coarser angular resolution 
will have D$_K \sim 0$. This is a conservative method for identifying blended
objects in $L'$, as the contrast between the brightest stars and the main 
body of fainter stars is greater in $L'$ than in $K$.

	Given the high stellar density near the center of M32, it is not surprising that
many of the objects have D$_K >> 0$, indicating that their 
photometric properties will likely be affected by blending at the angular 
resolution of the $L'$ data. An inspection of the images shows that stars 
with D$_K < 0.3$ tend to be in environments where stars of comparable brightness 
are not present within a few tenths of an arcsec; consequently, D$_K = 0.3$ was 
adopted as the threshold to distinguish between blended and unblended stars. 

	The $(L', K-L')$ CMDs of sources with D$_K < 0.3$ are shown in Figure 2. 
The data have been divided into three distance intervals, corresponding to 
projected distances $7.7 - 15.4$ pc (2 -- 4 arcsec), 
$15.4 - 28.5$ pc (4 -- 7.4 arcsec), and $28.5 +$ pc ($7.4+$ arcsec), 
as sample completeness varies with distance from the 
center of the galaxy. The angular intervals used to construct these CMDs 
are those that were adopted by Davidge et al. (2000). 
The projected distances assume a distance modulus of 24.5, which is based on an RGB-tip 
brightness of $I = 20.5$ (Freedman 1989; Davidge \& Jones 1992) and M$_I \sim 
-4$ for this feature (e.g. Lee, Freedman, \& Madore 1993). This distance modulus 
is consistent with that derived from an eclipsing binary in M31 (Ribas et al. 2005), which 
has the merit of being a primary distance indicator.

	Many of the brightest stars near the 
center of M32 are photometric variables (Davidge \& Rigaut 2004), and so the 
mixing of data recorded at different epochs smears color measurements. 
As the $K$ and $L'$ data were not recorded at the same epoch then 
the range of $K-L'$ colors seen in the CMDs is expected to be larger than 
the actual spread in intrinsic colors. This may explain, at least in part, 
why the AGB sequence in the CFHTIR data, which utilize measurements made at 
the same epoch and are discussed in \S 4, is better defined than that in Figure 2.

	The number of stars in the $7.7 - 15.4$ pc interval is modest, as the fraction 
of stars in this interval with D$_K > 0.3$ is much higher than at larger radii. 
Moreover, the vast majority of stars that are more than 28.5 pc from the nucleus 
have $L' > 14.4$, whereas at smaller radii there are stars as bright as $L' \sim 14$. 
Despite the procedure used to cull stars that are susceptible to blends, such a trend is 
suspiciously like that expected if the brightest stars at small radii are blends. 
Therefore, simulations were conducted to determine if residual 
blending has an impact on the photometry.

	Davidge (2001) co-added sub-fields with comparatively 
low stellar densities to simulate an environment with 
higher stellar density and assess the effect of blending near the center of M31; 
a comparison between the brightnesses of stars in the original and co-added datasets then 
provided a measure of the impact of crowding. This same procedure is employed here to 
determine if the brightest stars in the $15.4 - 28.5$ pc interval are blends.
We are not able to assess the impact of crowding at even smaller radii, 
as the NIRI dataset can not be divided into enough independent 
sub-fields to replicate the high central stellar density in M32.

	Three $5 \times 5$ arcsec$^2$ portions of the field near the 
edge of the $L'$ image were extracted and co-added. The 
$r-$band surface brightness measurements made by Kent (1987) indicate that the 
stellar density in the co-added frame is the same as that 20 pc ($\sim 5$ arcsec) from the 
nucleus. The brightnesses of individual stars were then measured in the co-added image.
The same sub-fields were extracted from the Hokupa'a $K-$band image 
and co-added. The result was smoothed to match the angular resolution of the 
$L'$ data, and D$_K$ was then computed for the various sources using the 
procedure described in \S 3.1. 

	The CMD of stars in the unstacked sub-fields is shown in the left hand panel 
of Figure 3. The CMD of sources with D$_K < 0.3$, which are 
those that the analysis in \S 3.1 suggests are likely not 
affected by blending, in the stacked images is shown in the middle panel of Figure 3. 
As might be expected given the small number of sources in the $15.4 - 28.5$ pc 
CMD in Figure 2, the number of objects identified as being unblended is modest, 
as many of the sources in the co-added field have been rejected as being susceptible 
to blending. Furthermore, it is evident that the 
photometric properties of individual objects with D$_K < 0.3$ 
{\it are} affected by blending, as the $K-L$ color of the brightest source is 0.8 mag 
different from its counterpart in the unstacked data. This is not unexpected, 
as even modest levels of crowding will introduce uncertainties into photometric 
measurements. Still, the photometry of stars with D$_K < 0.3$ is not 
systematically skewed by crowding. To demonstrate this, the brightnesses and colors 
of sources in the unstacked data were matched with 
their counterparts in the summed data, and the mean difference between the 
two sets of brightness and color measurements were computed. For stars with D$_K 
< 0.3$ we find that $<\Delta L> = -0.08 \pm 0.12$ magnitude, where the difference 
is in the sense stacked -- unstacked, while $<\Delta (K-L)> = -0.10 \pm 0.10$. 
That these differences do not differ significantly from zero indicates that the 
photometric measurements are not systematically skewed.

	The CMD of the co-added field without the rejection of stars with D$_K > 0.3$ 
is shown in the right hand panel of Figure 3. The effects of blending 
are clearly apparent when the CMD in the right hand panel is 
compared with those in the left hand and middle panels, as the number of 
objects with $L' < 14.5$ in the right hand panel is much larger than in the left hand 
panel. The brightnesses and colors of stars with D$_K > 0.3$ are also systematically 
skewed by crowding. For those stars with D$_K > 0.3$ we find that $<\Delta L> = -0.36 \pm 
0.10$ and $<\Delta (K-L)> = 0.24 \pm 0.13$, indicating that the mean brightness and 
color are affected by crowding.

	In summary, the simulation discussed above suggests that the 
rejection of stars that are susceptible to blending using the D$_K$ statistic 
produces a sample of objects with photometric properties that are not skewed 
by crowding. The brightest stars with $r > 4$ arcsec in Figure 2 are 
likely not unresolved blends of fainter stars. Thus, the brightest AGB stars near the 
center of M32 have $L' \sim 14.0$. 

\subsection{Comparisons with the Galactic Disk and Bulge}

	The $(M_{L'}, K-L')$ CMDs of stars with projected distances 
between 15.4 and 50 pc from the nucleus of M32 are shown in Figure 4. 
Also shown are data for (1) luminous M giants in the Galactic bulge from 
Frogel \& Whitford (1987), (2) LPVs near the Galactic Center from Wood, 
Habing, and McGregor (1998), and (3) luminous Galactic disk AGB stars that were studied
by Le Bertre (1992; 1993). The Le Bertre (1992; 1993) samples contain stars with redder 
$K-L$ colors and higher M$_L$ brightnesses than in the Frogel \& Whitford (1987) sample, 
although the two datasets appear to form a more-or-less continuous sequence 
on the CMD. While many of the stars in the Wood et al. (1998) sample also fall along 
the sequence defined by the Frogel \& Whitford (1987) M giants, there is a spray of 
objects that depart from this trend, and have $K-L$ colors comparable to those of the disk 
AGB stars.

	The stars in the three comparison datasets 
were selected at different wavelengths; the majority of stars in the Frogel \& Whitford 
sample are M giants selected from visible wavelength spectra (e.g. Blanco 1986), while 
the majority of the Le Bertre and Wood et al. samples were identified in the 
infrared. The majority of the stars discussed by Frogel \& Whitford (1987) have 
$V-K < 9$. While the $V-K$ colors of the objects with the reddest $K-L$ colors in the Le 
Bertre (1992; 1993) samples are not published, the fact that they are not listed 
in the USNO-A2.0 Catalogue (Monet et al. 1998) suggests that $V-K > 10$. 
Similar sources in Baade's Window would have $V > 17$ after taking into 
account foreground extinction, and so may have been missed 
in the photographic surveys discussed by Blanco (1986).

	The brightest AGB stars near the center of M32 fall along the sequences defined 
by Galactic bulge M giants and Galactic disk AGB stars in Figure 4. 
The Galactic disk and Galactic bulge sequences overlap
when M$_{L'} > -9$, and so some of the fainter stars near the 
center of M32 may be AGB stars like those in the Galactic bulge.
Still, when M$_{L'} < -9$, the M giants in the Frogel \& Whitford sample tend to have 
bluer $K-L'$ colors than the stars in M32, whereas many of the luminous Galactic 
disk AGB stars have $K-L'$ colors that overlap with 
those of the brightest stars in M32. The peak M$_{L'}$ 
of AGB stars in M32 is M$_{L'} \sim -10.4$, and this is in reasonable agreement with the 
peak brightness in the Galactic disk sample in Figure 4, although there are three stars 
with M$_{L'} < -12$ in the Le Bertre (1993) sample of oxygen-rich OH/IR stars that are 
not shown in Figure 4. 

	The Galactic disk AGB stars studied by Le Bertre (1992) are C stars, 
and the infrared SEDs of these objects are consistent 
with effective temperatures $\leq 1800$ K, with the majority having 
effective temperatures $< 1000$ K (Le Bertre 1997). That the brightest AGB 
stars in M32 have $K-L'$ colors like Milky-Way C stars is potentially of interest, as 
Davidge (1990) found that $60\%$ of the integrated light from M32 near $2\mu$m comes from 
AGB stars, and that $20\%$ of the AGB light may come from C stars. 
Still, the $H-K$ colors of stars on the upper AGB in M32 are 
consistent with these objects having SEDs indicative 
of M giants, rather than C stars (Davidge et al. 2000). The possible existence of C stars 
in M32 is addressed further in \S 4. 

\section{THE STELLAR CONTENT IN THE CFHTIR FIELD}

\subsection{Photometric Measurements}

	The $JHK$ brightnesses of stars in the CFHTIR data were measured with the 
PSF-fitting program ALLSTAR (Stetson \& Harris 1988), using co-ordinates and 
PSFs obtained from tasks in DAOPHOT (Stetson 1987). The photometric calibration 
was defined using standard stars from Hawarden et 
al. (2001), which were observed during the course of the three night observing run. 
The photometric calibration was checked against objects 
in the 2MASS point source catalogue (Cutri et al. 2003). These comparisons 
were restricted to stars with $K < 14.5$, which is the brightness range where 
the uncertainties in the 2MASS $K-$band measurements tend to be below $\pm 0.1$ mag. 
Moreover, to avoid the crowded main body of M32, these comparisons were 
further restricted to stars in the southern half of the CFHTIR field.
The differences in brightness, in the sense 2MASS -- CFHTIR, 
are $\Delta J = 0.00 \pm 0.02$, $\Delta H = -0.01 \pm 0.04$, and $\Delta K = 0.05 
\pm 0.02$, where the quoted uncertainties are the standard errors of the mean. 
The photometric calibration of the CFHTIR data is thus consistent with 
the 2MASS measurements.

	Completeness fractions and the random uncertainties in the photometric 
measurements were estimated by running artificial star experiments. 
The artificial stars were assigned colors that track the 
dominant red plume in the CMDs. There is a pronounced gradient in 
stellar density across the image, and so completeness and the random errors in the data 
vary with distance from the center of the galaxy. The completeness curves for 
stars that fall within projected major axis distances of 0.2 -- 
0.4 kpc and 0.8 -- 1.0 kpc, which are the inner and outer radial intervals considered 
for the photometric analysis, are compared in the upper panel of Figure 5. Following the 
procedure used to construct LFs in \S 4.2, artificial stars were 
considered to be recovered only if they were detected in both $H$ and $K'$.

	The completeness curves in Figure 5 show only modest differences 
due to stellar density. The completeness curve for the 0.8 -- 1.0 kpc 
interval falls systematically above that for the 0.2 -- 0.4 kpc interval, as 
expected given the difference in stellar density. Still, the brightnesses at which 
the completeness fraction is 50\% differ by only a few tenths of a magnitude.

	Crowding has an impact on the random uncertainties in the photometric 
measurements and the incidence of blending, in which star images merge and appear as a 
single object. While almost all stars in the CFHTIR data are likely blends, in the 
majority of cases a very bright star is blended with a very faint one, so that the impact 
on the photometric properties of the brighter star is minor. Only when two or more 
relativly bright stars merge together is the effect of blending significant 
in the present study.

	The artificial star data can be used to investigate the effects of 
blending, and the statistic of interest is the difference between the actual 
and recovered stellar brightnesses, $\Delta K$. The distribution of $\Delta K$ 
values at $K = 18$, which is near the brightness of the RGB-tip in 
M32 (Davidge 2000), for the 0.2 -- 0.4 kpc and 0.8 -- 1.0 kpc intervals 
are compared in the lower panel of Figure 5. While the two 
$\Delta K$ distributions both peak near $\Delta K = 0.1$, the distributions 
have very different widths. The $\Delta K$ distribution for the 0.2 -- 0.4 kpc 
interval is much broader than that at larger radii, with a tail extending out 
to $\Delta K = 0.6$. While the distribution for the 0.8 -- 1.0 kpc 
interval contains some stars with large $\Delta K$, they represent a much smaller 
fraction of the total population than at smaller radii. The absence of stars with 
$\Delta K > 0.7$ magnitude indicates that blending between stars of comparable brightness 
is likely not a major consideration at $K = 18$ in the portion of the CFHTIR data 
that samples the outer regions of M32. However, blending 
at this brightness is significant in the 0.2 -- 0.4 kpc interval. Consequently, 
the analysis of the CFHTIR data is restricted to stars with $K < 17$, which is 
well above the RGB-tip. The dispersion in $\Delta K$ at this brightness in the 
0.2 -- 0.4 kpc interval is roughly one-third that at $K = 18$, indicating that 
the effects of crowding are much reduced.

\subsection{The Photometric Properties of AGB Stars in the Outer Regions of M32}

	The $(K, H-K)$, $(K, J-K)$, and $(K, CO)$ CMDs of 
stars that are located between 0.2 -- 0.4 kpc (52 -- 104 arcsec), 
0.4 -- 0.6 kpc (104 -- 156 arcsec), 0.6 -- 0.8 kpc (156 -- 208 arcsec), 
and 0.8 -- 1.0 kpc (208 -- 260 arcsec) from the center of M32, where the boundaries 
are projected distances along the major axis assuming an ellipticity 
of 0.15 (Kent 1987) and a distance modulus of 24.5, are shown in Figures 6, 7, and 8. 
The properties of stars within 0.2 kpc of the nucleus are not considered, as crowding 
becomes much more of an issue in that portion of the galaxy at the angular resolution of 
the CFHTIR data. Following conventional practice, the CO index is a color measurement 
found by subtracting the magnitudes recorded through the CO and $K-$continuum filters. 
The various radial intervals into which the data have been sorted subtend comparably 
sized areas on the CFHTIR image, and so any differences in the number of 
objects on the CMDs are driven by changes in stellar density 
on the sky, rather than differences in angular coverage. 

	The RGB-tip occurs near $K \sim 17.8$ in M32 (Davidge 2000), and so the bright 
end of the red plume that dominates the CMDs in Figures 6 -- 8 is made up of AGB stars. 
There is also a smattering of stars with $K < 15$ in the two outer annuli. These 
objects have $J-K \sim 0.8$ and CO $\sim 0.2$, and we suspect that they are 
red supergiants in M31. These objects may be outlying members of stellar association 
A147, which lies to the south of M32, although the CFHTIR field does not overlap 
with the boundaries marked for A147 in the Hodge (1981) atlas.

	The fraction of stars in each annulus that are 
interlopers from M31 increases towards larger distances from the center of M32. 
However, using surface brightness measurements made from 
the 2MASS Nearby Galaxy Catalogue (Jarrett et al. 2003) $K-$band image of M31, 
it is found that nowhere in the CFHTIR field do stars belonging to M31 dominate. 
In the 0.8 -- 1.0 kpc interval, where the fractional contamination from M31 is largest, 
surface brightness measurements of M32 and the surrounding areas indicate that 
the expected ratio of M32 to M31 stars is at least 3:1. This ratio 
climbs quickly towards smaller galacocentric distances. In the 
0.6 -- 0.8 kpc interval the ratio of M32 to M31 stars is at least 5:1, while in the 
0.4 -- 0.6 kpc interval it is at least 9:1. There is a modest gradient in the 
density of M31 stars across the CFHTIR field, and this introduces a potential source of 
uncertainty in the amount of contamination from M31. However, the change in surface 
brightness between the northern and southern edges of the CFHTIR field is only 
$\sim 0.3$ mag arcsec$^{-2}$, and this does not have a major impact on the 
estimated level of contamination. In future studies it would be 
of interest to place tighter constraints on the number of stars that belong to M31 
by observing a control field that is well separated from M32. Kinematic 
measurements may also provide a means of identifying objects that do not belong to M32.

	While the AGB sequences in Figures 6 and 7 are relatively narrow, 
there is a population of objects to the right of the upper AGB. 
Carbon stars form a spray of objects redward 
of the M giant sequence on the $(K, H-K)$ and $(K, J-K)$ CMDs of moderately 
metal-poor systems with intermediate-age populations (e.g. Davidge 2003; 2005). 
Indeed, the majority of stars with $(J-K) > 1.6$ in the LMC are C stars (e.g. Hughes \& 
Wood 1990). Freedman (1992) found very red objects in the outer regions of M32, 
which she suggested may be C stars. Davidge (2000) noted that similar 
objects were not present in his data, although the modest field coverage meant that 
rare stars may have gone undetected. The current data cover a much larger field of view, 
and so can be used to re-visit the issue of very red stars in M32.

	The portions of the CMDs with $J-K > 1.6$ and M$_K < -7.25$, where the latter is 
the faint cut-off defined by Davidge (2005) for investigating C stars in the dE companions 
of M31, are indicated in Figure 7. The number of stars in the C star region increases 
towards smaller radii, suggesting that these objects belong to M32, rather than M31. 
The objects in the C star region of the CMDs have CO indices that are consistent 
with them being evolved, late-type stars. This is demonstrated in Figure 9, 
which shows the $(CO, J-K)$ two-color diagram (TCD) of objects with M$_K < -7.25$. 
The reddest stars have CO indices that stay roughly constant when $J-K > 1.6$, with 
$<CO> \sim 0.20 - 0.25$. The mean CO index of the very red stars also 
does not change with radius.

	In \S 3, we discussed a population of luminous red stars near 
the center of M32. The $K-L'$ colors of these objects are 
redder than those of optically selected M giants in the Galactic bulge, suggesting that 
they are embedded in dusty circumstellar envelopes. Are similar stars 
seen in the portion of M32 that was imaged with the CFHTIR? If these stars are 
present then they will appear as very red objects in the $(K, J-K)$ and $(K, H-K)$ CMDs, 
with $K-$band brightnesses that are well below the AGB-tip defined by bluer stars. 

	As it turns out, the reddest stars in the $(K, J-K)$ CMDs are probably {\it not} 
the counterparts of the stars with very red $K-L'$ colors seen near the center of M32. The 
stars with $K-L' < 1$ studied by Le Bertre (1992; 1993), which have colors and brightnesses 
that are similar to the most evolved M giants in the Galactic bulge, typically have 
$J-K \sim 2$, whereas the stars with $K-L' > 1$ typically have $J-K \sim 4$, with the 
reddest object having $J-K \sim 6$; these results hold for both the C stars 
studied by Le Bertre (1992) and the oxygen-rich stars studied by Le 
Bertre (1993). The majority of red stars in the CFHTIR data in Figure 
9 have $J-K < 2.5$, and so would be expected to have $K-L' \leq 1$. The failure to detect 
stars with $J-K > 3$ in the CFHTIR data does not mean that such very red stars are absent; 
rather, objects with very red $J-K$ colors are difficult to detect because they 
are faint, especially in $J$, and so there is a bias against detection. A 
wide-field survey of the outer regions of M32 in $L'$ will provide a means 
of determining if a population of very red stars like those detected near the 
center of M32 is present throughout the main body of the galaxy.

\subsection{A Search for Radial Trends}

\subsubsection{Isochrones and Predicted AGB Properties}

	The near-infrared photometric properties of AGB stars depend on age and 
metallicity, and these dependences are examined in Figure 10, where 
selected near-infrared isochrones from Girardi et al. (2002) are compared. 
These isochrones were constructed from the models described by Marigo \& Girardi (2001), 
and include a semi-analytical treatment of the thermally pulsing AGB. The models 
do not include circumstellar extinction, and so heavily obscured stars, 
such as those found near the center of M32 (\S 3), are not considered.

	A comparison with the $(K, J-K)$ CMD of the 0.2 -- 0.4 kpc interval in the right 
hand panel of Figure 10 indicates that the 1 Gyr model roughly matches the AGB-tip 
brightness in this portion of M32. Nevertheless, the agreement between the observations 
and models is far from perfect, as there is a $\sim 0.1$ magnitude offset in $J-K$. This 
is not a metallicity effect, as models with higher metallicity have $J-K$ colors that 
are only a few hundredths of a magnitude redder than the solar metallicity sequence. 
It is also worth noting that while the scatter in the $J-K$ colors of stars in the 
$0.2 - 0.4$ kpc interval appears to suggest that there may be a spread in age and/or 
metallicity, it is demonstrated below that this scatter is mainly due to 
observational errors.

	An important caveat when estimating age from the AGB-tip is that many of 
the brightest AGB stars in M32 are LPVs (Davidge \& Rigaut 2004), and this blurs the age 
sensitivity of this feature. The models predict that 
the AGB-tip brightness in $K$ will change by 
$\sim 0.3$ magnitude per 0.3 dex change in age 
between 1 and 8 Gyr. If the brightest AGB stars 
have $\pm 0.5$ magnitude photometric variations in $K$ (e.g. Davidge \& Rigaut 
2004), then the actual AGB-tip may occur as faint as $K \sim 16.3$, and thus correspond to 
an age $2 - 4$ Gyr. 

\subsubsection{The Colors of AGB Stars}

	The comparisons in the middle panel of Figure 10 indicate that 
while the brightness of the AGB-tip is only mildly sensitive to metallicity, the $J-K$ 
color changes by $\sim 0.1$ magnitudes when $\Delta [M/H] \sim 0.3$ dex. 
The mean $H-K$, $J-K$, and CO colors of stars with $K$ between 16.5 and 17.5 in 
each distance interval, computed by applying an iterative $2.5 \sigma$ rejection 
routine to suppress outliers, are listed in Table 1. The estimated uncertainties in the 
mean $J-K$ values are over an order of magnitude smaller than what is expected if 
metallicity changes by 0.3 dex; hence, these data should be sensitive to modest 
radial metallicity variations in the AGB population. It is also worth noting that the 
$J-K$ colors in Table 1 are considerably redder than the integrated $J-K$ colors that 
have been measured for M32, which fall between $J-K = 0.8$ and 0.9 (Frogel et al. 
1978; Peletier et al. 1993). This is not surprising, as the color listed in Table 1 is 
that of the brightest AGB stars, which will be redder than the integrated colors of the 
galaxy.

	There is no evidence for systematic radial trends in the mean broad-band colors 
in Table 1. The entries in the second column of Table 1 constitute a simple control 
measurement, as the $H-K$ color has only a modest sensitivity to age and metallicity 
when compared with colors that span a broader wavelength interval. 
It is thus reassuring that $<H-K>$ is constant with radius. 
$<J-K>$ also shows no evidence for radial trends, suggesting that if 
there is a metallicity gradient then it must be modest among AGB stars in the outer 
regions of M32. For comparison, $<CO>$ in the 
$0.4 - 0.6$ and $0.6 - 0.8$ kpc intervals differ at the $2.3 \sigma$ level, 
whereas $<CO>$ in the $0.2 - 0.4$ and $0.4 - 0.6$ intervals is 
$2.5 \sigma$ smaller than the mean of the outer two intervals. However, based on the 
individual $<CO>$ entries in Table 1, there is no
evidence for a systematic gradient in the CO index.

	An examination of the color distributions provides additional insights into 
radial trends in stellar content. In Figure 11 the $J-K$ color distributions of stars 
with $K$ between 16.5 and 17 in the 0.2 -- 0.4 kpc and 0.8 -- 1.0 kpc intervals, 
normalized to the number of stars in each interval, are compared. Also shown is the 
distribution defined by all stars in this brightness range in the full 0.2 -- 1.0 kpc 
interval. The range of stellar brightnesses used to construct these distributions is 
a compromise between the needs (1) to have a reasonable number of stars per distance 
interval, (2) to keep the random errors in the photometry modest in size, and (3) to 
include AGB stars that span a range of ages; in regard to the last point, the brightness 
interval used here includes stars with ages $\leq 8$ Gyr (e.g. Figure 10).

	The dotted line in each panel shows the distribution expected for a simple 
stellar system that is broadened only by random photometric errors, as predicted from 
the artificial star experiments. The effects of crowding are much less 
pronounced at this brightness than at $K = 18$ (\S 4.1), and the random errors 
in the photometry are only slightly larger in the 0.2 -- 0.4 kpc interval than in the 
0.8 -- 1.0 kpc interval. This being said, the $J-K$ distributions in both radial 
intervals are significantly wider than expected if random errors in the photometry were 
the only source of scatter, indicating that there is a real dispersion in the colors of 
the brightest AGB stars. 

	A comparison with the Girardi et al. (2002) isochrones indicates 
that the widths of the color distributions in Figure 11 
are consistent with a $\pm 0.3$ dex spread in metallicity. 
However, other factors may contribute to broadening the color distributions. 
A large fraction ($\sim 80\%$) of the AGB population near the center of M32 are 
LPVs (Davidge \& Rigaut 2004), and the color variations that occur throughout the 
light cycles of these objects can reproduce the observed dispersion along the 
$J-K$ axis. To demonstrate this point, we consider the $J-K$ colors of LPVs in the 
Sgr I field that were studied by Glass et al. (1995), and are based on 
observations made at a single epoch. In accordance with the range in 
intrinsic brightnesses used to construct the color curves in Figure 11, 
only those datapoints in the Sgr I dataset that have M$_K$ between --7.5 and --8.0
are considered.

	The LPVs in the Sgr I sample have a broad range of intrinsic 
$J-K$ colors, due in part to differences in line of sight reddening, 
metallicity, and/or age. As the goal of the current exercise is to investigate 
the color variations due to stellar variability, these star-to-star 
differences in mean color should be removed. This was done by computing a mean color 
for those LPVs with three or more measurements with M$_K$ between --7.5 and --8.0, 
and then calculating the difference between the individual measurements and 
the mean color, $\Delta (J-K)$. The standard deviation of 
all $\Delta (J-K)$ values is $\sigma_{LPV} = \pm 0.10$ magnitude, and this was 
adopted as the dispersion in $J-K$ due to variability.

	Knowing the color variation inherent to LPVs as they cycle through 
their photometric phases, then a model color distribution 
can be constructed. This model must also include 
the effects of observational errors and the presence of non-LPVs. 
The non-variable population was simply modeled as a gaussian 
with a width comparable to that expected from photometric errors. 
To account for observational errors in the LPV model, a gaussian with a standard deviation 
$\sigma_{LPV}$ was convolved with the photometric error 
distribution predicted by the artificial star experiments. 
The LPV and non-LPV components were then added together with various LPV fractions.

	The simulated color distributions are compared 
with the measured $J-K$ distributions in Figure 12. 
Davidge \& Rigaut (2004) found that 80\% of the bright AGB stars near the center 
of M32 are LPVs. While an 80\% LPV model provides a reasonable match 
to the $J-K$ distribution of stars in the 0.2 -- 0.4 kpc interval, it 
overestimates the width of the 0.8 -- 1.0 kpc distribution. A better match to the 0.8 -- 
1.0 kpc $J-K$ distribution is obtained if roughly 50\% of the stars are LPVs.
Radial differences in the LPV content notwithstanding, the comparisons in Figure 12 
indicate that the widths of the $J-K$ color distributions in Figure 11 may be 
dominated by stellar variability, rather than the presence of stars spanning 
a range of ages and/or metallicities. The stars on the upper AGB of M32 may then 
have only a very modest (i.e. $<< 0.3$ dex) dispersion in metallicity.

	The dispersion in the CO indices is dominated by random errors. This is 
demonstrated in Figure 13, where the distribution of CO indices in the 0.2 -- 0.4 and 
0.8 -- 1.0 kpc intervals are compared. The CO distributions in both intervals are 
consistent with a peak CO index near 0.3, and a tail of objects with higher than average 
CO indices. It is evident that the distributions in Figure 13 are dominated by the 
large random errors in the photometric measurements, and the annulus-to-annulus 
differences in mean CO index in Table 1 are likely not significant.

\subsubsection{The Relative Numbers of AGB Stars}

	The number density of bright AGB stars per unit integrated mass in composite 
stellar systems is an age diagnostic. In the present study 
surface brightness is used as a proxy for projected mass density. The $K$ LFs 
obtained from the $(K, H-K)$ CMDs of each distance interval, scaled to the number counts 
expected for a system with M$_{r} = -15$, based on the surface photometry from Kent 
(1987), and M$_K = -16$, based on the 2MASS Extended Source Catalog 
(Jarrett et al. 2003), are compared in Figures 14 and 15. The 2MASS surface 
brightness profiles of M32 become progressively noisier when $r > 150$ arcsec, and 
so the comparisons in Figure 15 do not include the two most distant annuli 
in the CFHTIR data.

	The integrated light measurements in the $r$ and $K$ filters are dominated by 
stars at different evolutionary stages. The integrated $r-$band light is dominated by 
stars on the sub-giant branch and near the MSTO, rather than the very 
bright AGB stars that we have resolved in M32. On the other hand, the integrated 
$K-$band light is dominated by highly evolved stars (e.g. Davidge 1990), and so the 
number of bright AGB stars would be expected to scale well with $K-$band brightness. 

	The comparisons in Figures 14 and 15 indicate that the number density of bright 
AGB stars in M32 stays remarkably constant out to galactocentric distances of at least 
1.0 kpc. This agreement extends into the central regions of M32, as the LF for stars 
with projectected distances $\sim 30 - 50$ parsecs from the center of M32 
agrees with the LF obtained from data with $r > 0.2$ kpc after correcting for 
differences in the integrated brightness. The comparisons in Figure 14 also indicate that 
the apparent decline in peak AGB brightness when $r > 0.6$ kpc from the 
nucleus is not significant, as the average number of stars brighter than 
M$_K = 16$ in the 0.6 -- 0.8 kpc and 0.8 -- 1.0 kpc intervals agrees with that 
expected from the counts in the 0.2 -- 1.0 kpc interval. Thus, there is no 
evidence that the peak $K-$band AGB-tip brightness drops with radius in M32; 
rather, the number counts along the upper AGB are consistent with no gradient in 
mean age. 

	We close this section by noting that Choi et al. (2002) found that the light 
profile of M32 deviates from an R$^{1/4}$ law at 150 arcsec, which corresponds to a 
projected distance $\sim 0.6$ kpc. This change in the light profile is accompanied 
by a change in the isophotal ellipticity. The CFHTIR data show no evidence for a change 
in the number density or photometric properties of the brightest AGB stars at this 
distance. The physical processes responsible for this structural change have evidently not 
affected the bright AGB content.

\section{COMPARISONS WITH STARS IN THE BULGE OF M31}

\subsection{Motivation, Field Selection, and Photometric Measurements}

	Davidge (2001) found that while the AGB sequences in M32 
and the inner bulge of M31 have similar peak $K-$band brightnesses, 
the number density of bright AGB stars in these systems differ, 
in the sense that M32 is deficient in stars below the AGB-tip when compared with the 
inner bulge of M31 after scaling to account for differences in the integrated brightnesses 
of the areas studied. An important caveat is that this comparison relies on data 
that sample very different environments, with the M31 observations 
sampling the innermost regions of the bulge, and the M32 observations sampling a 
field that is over 2 arcmin from the center of the galaxy. This difference in 
environment is potentially significant, as the central regions of many bulges harbour 
photometrically distinct nuclei (e.g. Carollo et al. 2002), and so the AGB content of the 
inner bulge of M31 may not be representative of the entire bulge. Thus, it is of 
interest to compare the stellar contents of M32 and the bulge of M31 using data that probe 
regions with similar surface brightness, and so a field in the outer bulge 
of M31 was observed with CFHTIR during the November 2002 observing run. 

	Crowding has had a significant impact on previous efforts to study the 
stellar content of the M31 bulge (e.g. discussion in Stephens 
et al. 2003), and the location of the M31 CFHTIR field was selected to be a 
compromise between the needs to sample (1) the bulge at a point where crowding should not 
be a factor among the brightest stars during good ground-based seeing conditions, 
and (2) an area that is dominated by bulge stars. 
The `small bulge' decomposition model of Kent (1989) indicates that the 
surface brightness of the bulge in the center of the CFHTIR field is 19 mag arcsec$^{-2}$ 
in the $r-$band, while that of the disk is 20.4 mag arcsec$^{-2}$; in other 
words, stars in the bulge outnumber those in the disk by $\sim 4:1$ in the CFHTIR 
pointing. Hence, the majority of stars in the CFHTIR dataset belong to the bulge of M31.

	The photometric analysis of the M31 CFHTIR data followed the 
procedures described in \S 4.1, including the use of artificial star experiments 
to assess random errors in the photometry and estimate 
completeness. The artificial star experiments indicate that 50\% 
completeness occurs when $K \sim 18$. This is brighter than 
in the M32 data (\S 4.1), and is a consequence of the relatively high stellar density 
throughout most (but not all -- see below) of the M31 CFHTIR field.

\subsection{Comparisons with M32}

	The M31 CFHTIR field was divided into three radial intervals, 
corresponding to projected distances along the minor axis of 0.5 -- 0.8 kpc (130 -- 
208 arcsec), 0.8 -- 1.1 kpc (208 -- 286 arcsec), and 1.1 -- 1.4 kpc (286 -- 364 
arcsec). The $(K, H-K)$, $(K, J-K)$, and $(K, CO)$ CMDs of 
sources in these intervals are shown in Figures 16, 17, and 
18. The artificial star experiments predict that some of the brightest stars in the 0.5 
-- 0.8 kpc CMDs may be blends. To avoid potential problems due to blending 
and focus on a region that has an integrated $r-$band surface brightness that 
is comparable to that in the $0.2 - 0.4$ kpc interval in M32, 
only the data in the 1.1 -- 1.4 kpc interval is considered 
further. The brightest AGB stars in this portion 
of the M31 CFHTIR field have $K \sim 15.7$, which is within a few tenths of 
a magnitude of the peak AGB brightness near the center of M31 (Davidge 2001) and in M32.

	The $J-K$ and CO color distributions of stars with 
$K$ between 16.5 and 17 in the 1.1 -- 1.4 kpc interval are shown in Figure 19. 
The mean $J-K$ color of stars in the M31 field is comparable to what is seen in M32. 
Moreover, as was the case in the $0.2 - 0.4$ kpc interval in M32, the 
$J-K$ distribution can be matched by a model that combines 
random photometric errors estimated from artificial star experiments 
with a population of AGB stars in which 80\% are LPVs. 

	The CO index in the lower panel of Figure 19 shows 
the tail of high-CO index stars that was also seen in the M32 CO distribution. 
In fact, the CO distributions of the two galaxies are very similar.
This is demonstrated in the lower panel of Figure 19, where 
the dotted line is the CO distribution of stars in M32 in the 0.2 -- 0.4 kpc interval, but
shifted along the horizontal axis so that the peak matches the peak of the M31 
distribution. The shift applied to these data is within the estimated uncertainties in the 
photometric calibration; therefore, the need to shift the two distributions is likely 
not due to a real difference in the mean CO indices of the two systems, but is a 
consequence of uncertainties in the calibration.

	The $K$ LF of stars in the 1.1 -- 1.4 kpc distance interval in 
M31 is compared in Figure 20 with the LF of stars in the 0.2 -- 0.4 kpc interval in M32. 
The LFs have been normalized so that the star counts 
correspond to M$_{r} = -15$ (top panel) and M$_K = -18$ (lower 
panel). The normalizations were done using surface brightness measurements from 
Kent (1987) and Jarrett et al. (2003). The LFs are offset by a 
significant amount along the vertical axes, with the difference growing towards 
brighter magnitudes. These comparisons indicate that the outer regions of M32 have a 
higher density of bright AGB stars per integrated brightness than the outer bulge of M31. 
Such a difference would occur if M32 and the outer bulge of M31 have different 
mean ages, and this possibility is discussed further in \S 6.

\section{DISCUSSION \& SUMMARY}

	Broad- and narrow-band images spanning the $1 - 4\mu$m wavelength interval 
have been used to probe the bright AGB content of M32. Data recorded in $L'$ 
with NIRI on Gemini North are used to conduct the first study of 
individual stars in M32 at wavelengths longward of $2.5\mu$m. 
The $L'$ data cover projected distances out to $\sim 50$ parsecs from 
the galaxy center, and provide a means of identifying the most evolved AGB stars, which may 
be difficult to detect at visible and near-infrared wavelengths if they are embedded in 
dusty circumstellar envelopes. Images recorded through $J, H, K',$ and narrow-band filters 
of a field immediately south of the M32 nucleus with the CFHTIR sample roughly 25\% of the 
galaxy at projected major axis distances between 0.2 and 1.0 kpc. These data 
are used to investigate the radial properties of the bright stellar content in M32.

	A prime motivation for the study of the resolved stellar content of 
nearby galaxies is that direct comparisons can be made with predictions from 
the analysis of integrated light spectra. The three main results of this paper that 
relate to this motivation are as follows. First, there are 
luminous AGB stars near the center of M32 with $2 - 4\mu$m photometric properties 
that are similar to those of the most luminous AGB stars in the disk of the Milky-Way; 
these objects are likely intermediate-age AGB stars. Second, the bright 
AGB content of M32 in the $K-$band is well mixed throughout the main body of the galaxy. 
This result, which confirms earlier findings that were based on data covering much smaller 
areas, is contrary to what would be expected if there were a radial age gradient 
throughout the main body of the galaxy. Third, while the AGB sequences in M32 and the 
outer bulge of M31 have similar $K-$band peak brightnesses, and possibly even 
similar LPV fractions, the number of bright AGB stars per unit 
integrated light is higher in M32 than in the outer bulge of M31; this is indicative 
of a difference in stellar content. These results are discussed in the remainder of this 
section.

\subsection{An Intermediate Age Component in M32}

	The integrated light spectrum of M32 at visible wavelengths contains 
signatures of an intermediate age population (e.g. Worthey 2004; Rose et al. 2005 and 
references therein), and the properties of the brightest resolved stars in 
M32 verify that such a component is present. Spectra taken at visible wavelengths 
provide important leverage for measuring age because stars near the MSTO contribute 
significantly to the integrated light at these wavelengths. One of the pieces of 
information that can be obtained from the analysis of integrated light spectra at 
near-infrared wavelengths is the contribution that AGB stars make 
to the total system light, which is a function of age. Models 
generated by Maraston (1998) indicate that the AGB contribution in the $K-$band 
peaks near an age of 1 Gyr, where it can produce 80\% of the total light. 
The predicted AGB contribution drops to 50\% for an age of 3 Gyr, and 10\% for 
an age 6 Gyr. Davidge (1990) modeled the integrated spectrum of the 
center of M32 in the $1.5 - 2.1\mu$m interval and 
found that $\sim 60\%$ of the integrated $K-$band light comes from 
AGB stars. This result, which is evidence for an intermediate age population,
can be checked directly using the number counts of AGB stars measured from the CFHTIR data.

	The total light from AGB stars was found by integrating the $K-$band 
LF above the RGB-tip and comparing the result with the integrated 
$K-$band light in the area surveyed, which was computed from 2MASS data 
(Jarrett et al. 2003). Assuming that (1) the RGB-tip occurs at $K = 17.8$ (Davidge 
2000), and (2) the AGB can be modeled as a power law, based on the LF entries between 
$K = 15.5$ and 17.0, which is the brightness at which incompleteness sets in at smaller 
radii (\S 4), then we find that the AGB accounts for $70^{+30\%}_{-20\%}$ of the 
total $K-$band light. This agrees with the AGB contribution computed by Davidge (1990), 
and is indicative of a relatively young photometrically-weighted age for M32. 
The solar metallicity models generated by Maraston (1998) predict that this 
AGB contribution is indicative of ages in the range log(t$_{yr}) \sim 8.5 - 9.5$. When 
combined with the additional constraint that there are no 
MSTO stars with ages $< 1$ Gyr (Worthey et al. 2004), then 
a tighter age range log(t$_{yr}) \sim 9.0 - 9.5$ (i.e. $1 - 3$ Gyr) results.

	Stars near the AGB-tip are potentially important age probes, as 
they stand out with respect to the fainter, but more 
numerous, body of stars in a galaxy. The AGB-tip 
can be studied in areas where fainter age diagnostics, such as the morphology of 
the horizontal branch or the brightness of the MSTO, can not be 
detected. Complicating factors when using the AGB-tip brightness as an age estimator are 
that (1) the brightest AGB stars are relatively rare, and so a large area must be sampled 
to obtain a representative census -- this is likely the cause of the apparent 
drop in AGB-tip brightness in the CFHTIR CMDs when $r > 0.6$ kpc (\S 4), and (2) 
a large fraction of the brightest AGB stars are LPVs with amplitudes approaching 
a magnitude in $K$. Both of these factors blur the 
ability to measure the AGB-tip, and hence deduce ages. Davidge (2000) 
compared the near-infrared CMDs of the brightest AGB stars 
in M32 with theoretical isochrones and concluded that these 
objects have an age of a few Gyr. This is consistent with comparisons 
that are made with isochrones in \S 4. If photometric variability is such that 
the brightest stars at the peak of their light curves appear $\sim 0.5$ magnitudes in $K$ 
above the AGB-tip of non-variable sources, then comparisons with isochrones 
indicate that this may cause ages to be underestimated by a few Gyr.

	Mass that is lost from AGB stars may accumulate in circumstellar envelopes, 
and the extinction that can result from dust in these envelopes may also affect age 
estimates, as the most luminous AGB stars may be missed in surveys conducted at wavelengths 
where dust absorption is significant. The amount of 
material in envelopes is expected to depend on the mass of the 
progenitor, in the sense that envelope mass will increase with progenitor 
mass. Circumstellar dust will attenuate visible/near-infrared light from 
the star, and AGB stars in thick envelopes will thus appear as heavily reddened sources 
with a dominant thermal emission spectrum at wavelengths longward of $2.5\mu$m 
from heated dust grains.

	In \S 3 it was shown that the central regions of 
M32 contain a population of stars with $K-L'$ colors and peak $L'$ brightnesses 
that are similar to those of the brightest AGB stars in the disk of the Milky-Way. 
The photometric properties of the majority of $L'-$bright AGB stars in M32 
differ from those of the brightest AGB stars in the Galactic bulge and the bulge of M31 
(Davidge et al. 2006), in that they have higher $L'$ brightnesses and 
redder $K-L'$ colors. The $L'-$bright objects 
detected in M32 are almost certainly intermediate-age AGB stars.

	The $K-L'$ colors of the $L'-$bright AGB stars in M32 
suggest that they are embedded in dusty shells, and so 
are subject to heavy circumstellar extinction. Therefore, they will appear 
in near-infrared CMDs as red objects, likely falling below the AGB-tip defined by lower 
mass, but less reddened, objects. A population of red objects below the AGB-tip and to 
the right of the AGB sequence is seen in the CFHTIR $(K, J-K)$ CMDs, although it was 
argued in \S 4 that these objects are likely not the counterparts of the luminous AGB stars 
seen near the center of M32, as their $J-K$ colors are too small. 
It was further noted in \S 4 that luminous AGB stars like those found near the 
center of M32 may be too faint to detect with the CFHTIR 
observations. An imaging survey of the outer regions of M32 in $L'$ should
reveal if objects like those found near the center of M32 are also present at large radii.

	C stars are among the most conspicuous signatures of an intermediate-age 
population. Surveys of C stars in nearby galaxies suggest that the C/M ratio is 
a function of metallicity (e.g. Battinelli \& Demers 2005 and references therein). 
This trend is reproduced by models that track evolution on the thermally-pulsing 
AGB (e.g. Mouhcine \& Lan\c{c}on 2003a), and is due to 
a lower efficiency for C star formation as metallicity grows. The Mouhcine \& Lan\c{c}on 
(2003a) models predict that C stars will form in a Z=0.02 system, although the C/M ratio 
will be $2 - 10 \times$ lower than in a Z=0.008 system with the same age. Therefore, 
despite a relatively high mean metallicity, C stars might be 
expected in M32 if the progenitor population has a suitable age and size.

	Davidge (1990) found that the best agreement between the modeled and 
observed near-infrared spectrum of M32 was obtained if $20\%$ of the AGB light comes from 
C stars, as opposed to an AGB component that consists entirely of M giants. 
While conspicuous signatures of C stars, such as the Ballick-Ramsey 
C$_2$ band at $1.77\mu$m, are not evident in the composite M32 $H-$band spectrum 
obtained by Davidge (1990), these may be difficult to detect if blended with 
other molecular features, at least at moderately low spectral resolutions. The 
C star contribution predicted by Davidge (1990) is not 
greatly different from what is found in intermediate age LMC clusters with mean 
metallicities that are not too much lower than that inferred for M32 (Maraston 1998).

	The evidence for C stars in the resolved stellar content of M32 is more tenuous. 
In \S 3 it was argued that while the brightest stars in $L'$ near the center of M32 
have $K-L'$ colors that are similar to those of C stars in the disk of the Milky-Way, 
they have $H-K$ colors suggesting that they are obscured M giants. The nature of these 
stars could be investigated further with moderate-resolution spectra spanning the 
$3 - 4\mu$m wavelength interval. C stars have a prominent absorption feature due to 
C$_2$H$_2 +$ HCN at $3.1\mu$m (Ridgeway, Carbon, \& Hall 1978) that is seen even in 
highly reddened C stars (e.g. Le Bertre et al. 2005), although circumstellar 
emission may cause these features to be veiled (Matsuura et al. 2005). For 
comparison, the $3\mu$m spectra of moderately metal-poor M giants are almost featureless 
(Matsuura et al. 2005). While well known spectral signatures that can be used to 
distinguish between C and M stars are present at shorter wavelengths, it 
may prove difficult to obtain spectra of the objects near the center of M32 
at wavelengths shortward of $3\mu$m because the targets are fainter at these wavelengths, 
and are more susceptible to blending with unreddened stars.

\subsection{Is There An Age Gradient in M32?}

	Based on the strengths of absorption features in the integrated visible 
light spectrum of M32, Worthey (2004) and Rose et al. (2005) conclude that 
mean age and metallicity vary with radius, in the sense that older, more metal-poor 
populations occur at larger radii. In the current paper, three properties of AGB 
stars -- their peak brightness, their relative numbers per unit integrated brightness, 
and the histogram distribution of their broad-band colors -- are considered 
together to assess if age changes with radius in M32. 
The data used by Davidge et al. (2000), which resolve the brightest stars within a few 
arcsec of the galaxy center, have been used to bridge the spatial coverage of the 
spectroscopic and CFHTIR datasets. The results suggest that age does not vary with 
radius, and that the population traced by the brightest AGB stars 
are very well mixed throughout the galaxy. Lacking observations of RGB stars, 
little can be said about a metallicity gradient in M32 based on the present data, save 
that the brightest AGB stars appear to have the same metallicity at all radii. 

	The brightest AGB stars in M32 have $K \sim 15.5$, and stars of this brightness 
are seen out to projected distances along the semi-major axis of at least 0.6 kpc. 
While the peak brightness of the CMDs appear to drop when $r > 0.6$ kpc, 
this is likely a consequence of small number statistics (\S 4). 
The presence of AGB stars with the same peak brightness does not in itself 
argue against an age gradient, as the number of these objects per unit integrated mass, 
and hence the mean age, could still change with radius. Rather, when considered 
on its own, a constant peak brightness near $K \sim 15.5$ indicates only that an 
intermediate age population is present throughout M32.

	The relative number of AGB stars per unit integrated mass 
is a robust means of determining if there is a radial age gradient.
Consider a hypothetical system that is made up of two populations, one of 
which is `old', containing an AGB component that is not much brighter than the RGB-tip, 
and the other `young', containing an extended bright AGB sequence. If 
the younger population is more centrally concentrated than the older 
population then mean age grows with increasing distance from the center 
of the system, and the number density of the brightest 
AGB stars with respect to the total mass of stars in a given radial interval 
drops with increasing radius. 

	While the ratio of AGB stars to total mass is 
an age diagnostic, in the absence of suitable dynamical measurements 
then one is forced to use integrated brightness as a proxy for 
mass, as has been done in Figures 14 and 15. Complications arise because the M/L 
ratio of an integrated system depends on its mean age, in the 
sense that the M/L ratio becomes smaller towards progressively younger ages. A 
change in M/L ratio due to age may compensate to some extent for a drop in the number 
density of bright AGB stars with respect to fainter objects.

	The issue of the age-dependence of the M/L ratio notwithstanding, 
if there were an age gradient in M32 then the good agreement between the LFs 
shown in Figure 14 would require that age change radially in such a 
finely tuned way that the M/L ratio compensates for any differences in the 
fractional contribution that the brightest AGB stars make to the total light.
The uncertainty due to the age dependence of the M/L ratio can be further mitigated 
by normalizing the LFs to integrated brightnesses in different filters, as the 
M/L sensitivity to age is wavelength dependent. For example, the 
models considered by Mouhcine and Lan\c{c}on (2003b) indicate that the M/L ratio in $V$ 
changes by 0.9 dex when age changes from 1 Gyr to 10 Gyr, while the M/L ratio in $K$ 
changes by only 0.5 dex over the same age range. It is then significant that the 
number of AGB stars per unit $K-$band light, shown in Figure 15, does not change with 
radius in M32. When considered together, the results in Figures 14 and 15 then 
indicate that mean age likely does {\it not} increase towards 
larger galactocentric distances in M32, contrary to what 
was found by Worthey (2004) and Rose et al. (2005). 

	The color distribution of stars in a given brightness interval is also 
sensitive to the dispersion in age and metallicity. In Figure 12 it was demonstrated that 
the spread in the $J-K$ colors of bright AGB stars in M32 
can be explained by a combination of photometric errors and the changes in color 
experienced by LPVs as they cycle through their light curves. That these two effects 
can largely account for the observed dispersion in $J-K$ color suggests that 
the brightest AGB stars in M32 likely have only a modest range in age 
and metallicity. The mean $J-K$ color of bright AGB stars does not change 
with radius, suggesting that the mean metallicity of the bright AGB component 
also does not change with radius.

	The shape of the $J-K$ distribution may change with radius, 
in the sense that the number of stars that are LPVs and/or the amplitude range 
of LPVs may drop with radius; as discussed in \S 4, the fractional LPV content in the 
$0.8 - 1.0$ kpc interval may be $\sim 30\%$ smaller than that in the $0.2 - 0.4$ kpc 
interval. A trend of decreasing LPV content with increasing 
radius is seen in NGC 5128 (Rejkuba et al. 2003). The amplitude and period distributions 
of LPVs will provide additional insights into the stellar content of the outer 
regions of M32. While a determination of the period distribution will require multi-epoch 
observations, the fraction of LPVs and their amplitude distributions 
could be probed directly with only a single second epoch observation of the CFHTIR 
field by applying the procedure described by Davidge 
\& Rigaut (2004) to determine the fraction of LPVs as a function of radius.

	The CO index, which measures the strength of the first overtone CO 
(2-0) bandhead, is used in the current paper as an additional probe of the nature 
of the brightest AGB stars in M32. The brightest AGB stars are found to have 
uniform CO strengths throughout M32, further reinforcing the notion that these stars 
come from a population that is well mixed throughout the galaxy. 
The integrated $K-$band light from M32 is dominated by 
AGB stars (e.g. Davidge 1990 and \S 6.1), and so the absence of a gradient in the CO 
indices of the brightest AGB stars is consistent with the flat CO color profile 
seen in the integrated light of this galaxy (Peletier 1993).

	Based on the properties of the brightest AGB stars, we conclude 
that there is no evidence for a radial age gradient in M32. 
While the results of this paper challenge the conclusion reached by Worthey (2004) and 
Rose et al. (2005) that mean age changes with radius in M32, population gradients 
{\it do} occur in this galaxy, as the strengths of absorption features in 
the integrated spectrum change with radius (e.g. Worthey 2004; Rose et al. 2005). 
There is also evidence for a mild UV color gradient in M32, that is reminiscent 
of what is seen in classical ellipticals (e.g. de Paz et 
al. 2005). We note in passing that since there is no evidence for a gradient 
in the properties of the brightest AGB stars then the cause of the UV 
gradient is likely not tied to the brightest AGB stars or their descendants. 
We emphasize that the CFHTIR data do not place firm limits on the radial variation 
of mean metallicity, as the RGB is not sampled. It thus remains to be determined if 
metallicity is the prime driver behind the population gradients in M32.

\subsection{Comparing the Stellar Contents of M32 and the Bulge of M31}

	Bica et al. (1990) compare the stellar contents of M32 and the bulge of M31 
using spectra of both galaxies that were obtained with the 
same instrument and were analysed using the same technique. 
They find that a range of ages are present in the central 
regions of M31, and that while the dominant component is old and metal-rich, 
a very young component is also present. While 
a component with properties that are similar to the intermediate 
age population in M32 is present in M31, its contribution to the integrated 
$V-$band light is only one half what it is in M32. The analysis of these spectra 
thus predict that M32 has a younger mean age than the bulge of M31. 

	How does this compare with the resolved stellar content? We 
find that the most luminous AGB stars near the center of M32 have M$_{L'}$ and 
$K-L'$ colors that are similar to the brightest AGB stars in the disk of the Milky-Way, 
while the most luminous AGB stars near the center of M31 have M$_{L'}$ and $K-L'$ 
that are similar to stars in the Galactic bulge (Davidge et al. 2006). This suggests that 
the center of M32 contains stars that are younger than those near the center of M31. 
This does not agree with the analysis of the Bica et al. (1990) spectra, which 
predicts that there is a very young component near the center of M31 but not in M32. 
Moreover, working in the $K-$band, Davidge (2001) found a higher number density 
of bright AGB stars in the inner bulge of M31 than in the outer regions of M32, 
which is suggestive of a centrally-concentrated intermediate-age 
population that is larger in size in M31 than in M32. The 
seemingly contradictory results that are drawn from the $L'$ measurements in \S 3 
and the comparison conducted by Davidge (2001) can be reconciled if (1) the 
center of M32 contains younger stars than in M31 but (2) the 
overall density of stars that formed during intermediate epochs is smaller in M32 
than in the inner bulge of M31. Such a star-forming history is not consistent with the 
Bica et al. (1990) results.

	There are absorption line gradients in the bulge of M31 that are 
indicative of radial changes in both age and metallicity (e.g. Davidge 1997), and 
Puzia, Perrett, \& Bridges (2005) find that the bulge of M31 has spectroscopic 
characteristics at large radii that are indicative of an old system, with an age that is 
comparable to that of the oldest M31 globular clusters. The comparisons in Figure 20 
indicate that the outer bulge of M31 is deficient in 
stars below the AGB-tip when compared with M32. This 
deficiency occurs when the data are normalized both in $r$ and $K$, and so is 
not a consequence of problems with the surface photometry. The comparisons in Figure 
20 are suggestive of a difference in mean age, although M32 and the bulge of M31 
may contain stars spanning a similar range of ages. More specifically, while the 
two systems appear to have different peak $K$ brightnesses, the number of bright AGB 
stars is much lower in the outer bulge of M31 than in M32, and so 
stochastic effects may cause the peak brightness in M31 to be 
underestimated. Thus, the outer bulge of M31 may contain a component with an age 
that is similar to that of the brightest AGB stars in M32.

	The comparisons in Figure 20 suggest that M32 and the outer bulge of M31 
have had different star forming histories. The 
two LFs become significantly different near $K \sim 16.8$, which corresponds to 
M$_K = -7.7$. An AGB star in a solar metallicity system with an age $\sim 7$ Gyr would have 
such a peak brightness, although there is uncertainty introduced by photometric 
variability among AGB stars. The uncertainty in the relation between peak AGB brightness and 
age notwithstanding, it appears that the star forming histories of M32 and the outer 
bulge of M31 have differed over a significant fraction of their lifetimes.

\subsection{Implications for the Evolution of M32}

	The radial distribution of stars in galaxies provides insight into the 
past evolution of these systems. A remarkable property of M32 is that the brightest AGB 
stars are mixed uniformly throughout the galaxy, and this 
must be explained by any model of its past history. 
It has been suggested that M32 may have once been a disk system 
that was disrupted by interactions with M31 (Bekki et al. 2001; 
Graham 2002). In the context of such a model the bright AGB stars may be the remnant of 
the last burst of star formation in a now defunct disk around M32, while the 
older substrate is the remnant bulge. Although involving a more extreme 
case, simulations that explore the interaction between a supposed progenitor of 
$\omega$ Cen and the Galaxy support the feasibility of this interpretation, and indicate 
that tidal interactions can mix stars throughout the smaller system. 
Simulations indicate that the core of the $\omega$ Cen progenitor can remain intact, while 
the surrounding disk is largely disrupted, but not before being tidally stirred (e.g. 
Bekki \& Freeman 2003). While such a process may mix a recently formed 
population uniformly throughout M32, it might also be expected to 
mix populations that were already in place, and thereby flatten any metallicity gradients 
that may have been present in the bulge of the M32 progenitor. If such mixing did occur 
then parameters other than mean age and mean metallicity may be responsible for 
the absorption line gradients in M32.

	If there have been major interactions between M31 and M32 then one might 
expect to see material stripped from M32 that is being assimilated by M31. Ferguson et 
al. (2002) discuss reasons why M32 might be the origin of at least some of the 
material in streams detected around M31, although there are potential kinematic 
difficulties with M32 being the origin of the giant stream (Ibata et al. 
2004). The material stripped from M32 may not be 
restricted to individual stars, and it is worth noting in this regard that 
M32 is completely devoid of globular clusters. Beasley et al. (2005) and Puzia et al. 
(2005) find that some M31 globular clusters have ages 
that overlap with the youngest populations in M32, and these clusters 
appear to share a common evolutionary heritage with M32 that is distinct from 
that of M31. In particular, the chemical mixtures in M32 and the intermediate age 
M31 clusters are similar, and the integrated light spectra of the clusters lack the 
strong CN bands that are the hallmark of other M31 
clusters and the bulge of M31, but are absent in the integrated spectrum of M32.
Comparisons with nearby spiral galaxies suggests that strong CN absorption, 
which may be due to an overabundance of nitrogen, is a unique characteristic 
of the M31 globular cluster system (Puzia et al. 2005). 

	Beasley et al. (2005) argue that the intermediate age M31 clusters are too 
metal-poor to be associated with M32, and suggest that they may have originated 
in NGC 205. However, Puzia et al. (2005) find intermediate age clusters that are 
moderately metal-rich. The C star content of NGC 205 also indicates that it likely 
did not experience a large burst of star formation during the past few Gyr 
(Davidge 2005). Given the evidence at hand, it may be premature to reject 
the possibility that the intermediate age clusters in 
M31 either formed in M32 or from gas and dust that was stripped from M32.

\parindent=0.0cm

\clearpage

\begin{table*}
\begin{center}
\begin{tabular}{cccc}
\tableline\tableline
Major Axis & & & \\
Distance (kpc) & $<H-K>$ & $<J-K>$ & $<CO>$ \\
\tableline
0.2 -- 0.4 & $0.501 \pm 0.007$ & $1.452 \pm 0.005$ & $0.351 \pm 0.009$ \\
0.4 -- 0.6 & $0.488 \pm 0.009$ & $1.450 \pm 0.006$ & $0.329 \pm 0.011$ \\
0.6 -- 0.8 & $0.506 \pm 0.010$ & $1.410 \pm 0.008$ & $0.384 \pm 0.021$ \\
0.8 -- 1.0 & $0.478 \pm 0.011$ & $1.434 \pm 0.010$ & $0.370 \pm 0.016$ \\
\tableline
\end{tabular}
\end{center}
\caption{Mean Colors}
\end{table*}

\clearpage

\begin{figure}
\figurenum{1}
\epsscale{0.75}
\plotone{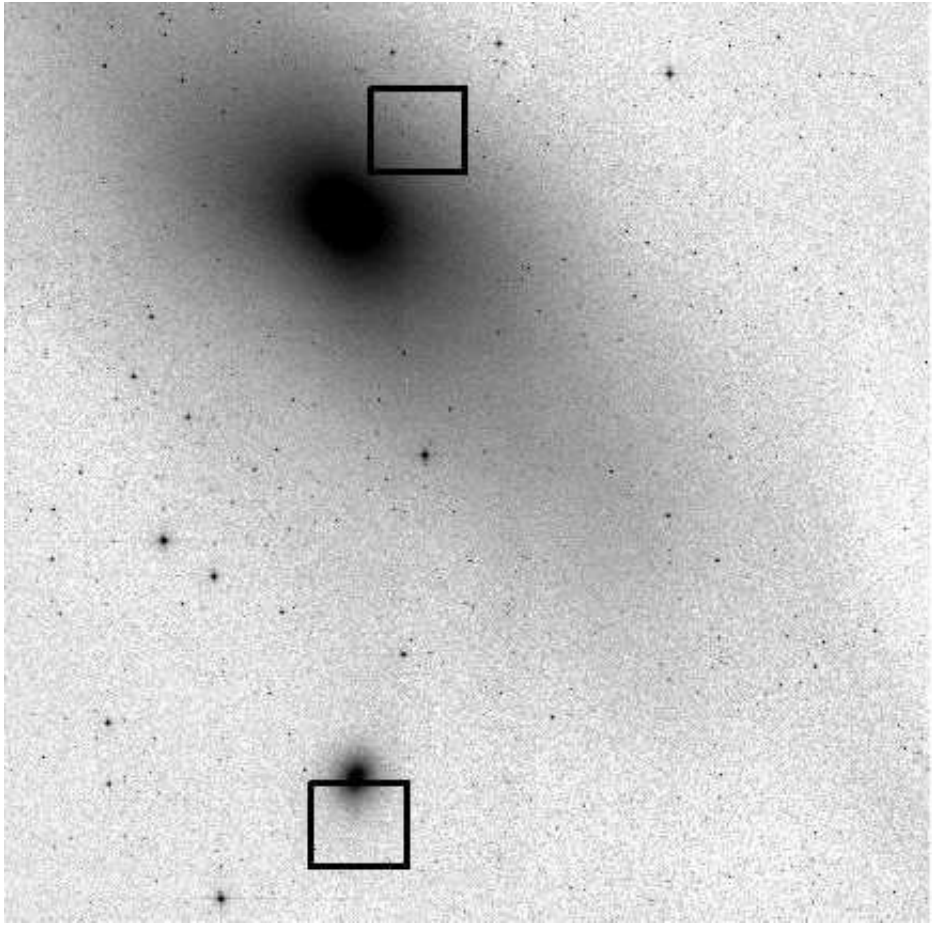}
\caption
{The locations of the M32 and M31 CFHTIR fields are shown on this $40 \times 40$ arcmin$^2$ 
section extracted from the 2MASS Large Galaxy Atlas (Jarrett et al. 2003) 
$K-$band image of M31. North is at the top, and East is to the left.} 
\end{figure}

\clearpage

\begin{figure}
\figurenum{2}
\epsscale{0.75}
\plotone{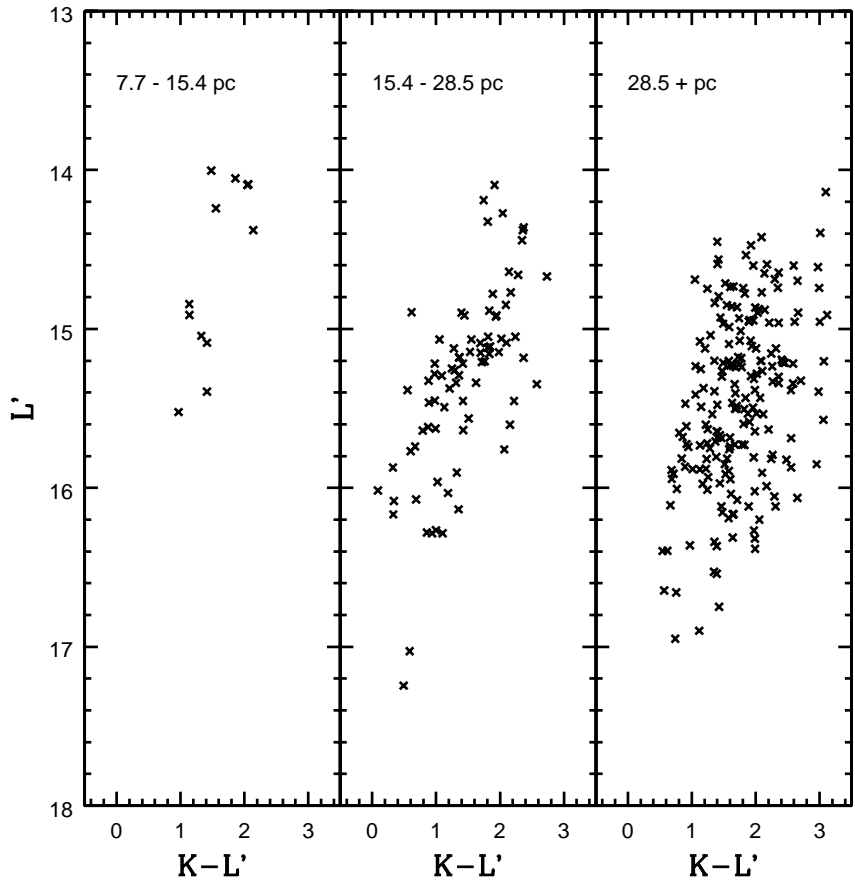}
\caption
{The $(L', K-L')$ CMDs of stars in three radial intervals near the center of M32. 
The projected distances are measured from the center of the galaxy 
and assume a distance modulus of 24.5. Only stars that are likely 
not blends, with D$_K < 0.3$ (see text), are shown.}
\end{figure}

\clearpage

\begin{figure}
\figurenum{3}
\epsscale{0.75}
\plotone{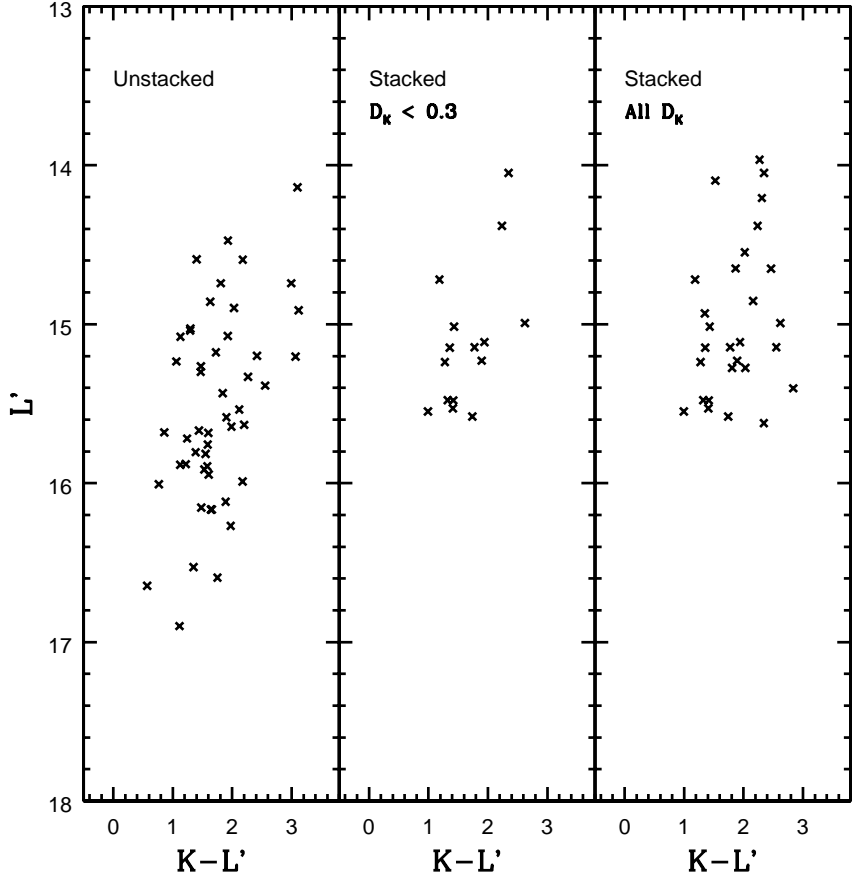}
\caption
{The results of simulations that investigate the impact of crowding on the $L'$ data. The 
left hand panel shows the CMDs of stars in three $5 \times 5$ arcsec$^2$ sub-fields 
near the edge of the NIRI field. If summed, these sub-fields have an $r-$band surface 
brightness that is comparable to that in M32 at a projected distance of 20 pc (5 arcsec) 
from the nucleus. The middle panel shows the CMD constructed from the summed 
sub-fields, which was photometered in the same way as the initial 
data, including the rejection of stars that are likely blends. 
The right hand panel shows the CMD without applying the D$_K$ criterion to 
cull blended objects. Note that (1) many of the stars in the co-added 
field are rejected as blends, and (2) the number of stars with $L' < 14.5$ 
in the middle panel is not markedly different from that in the 
left hand panel.}
\end{figure}

\clearpage

\begin{figure}
\figurenum{4}
\epsscale{0.75}
\plotone{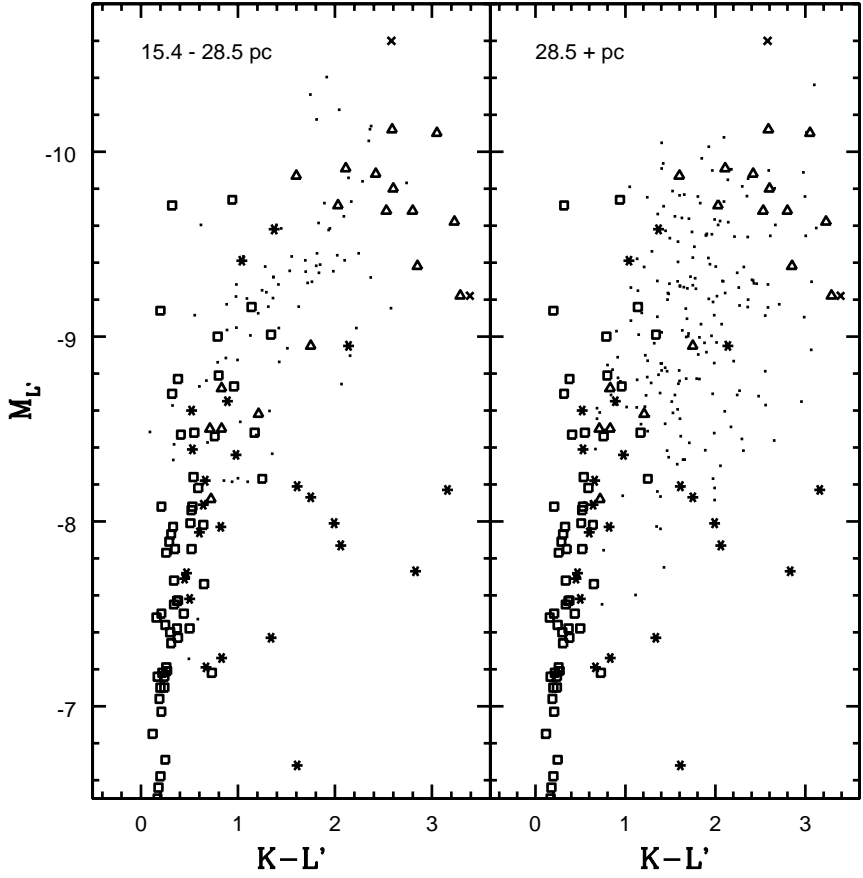}
\caption
{The $(M_{L'}, K-L')$ CMDs of stars in two radial intervals near the center of M32. 
A distance modulus of 24.5 has been assumed, based on the $I-$band brightness 
of the RGB-tip. Stars in M32 are plotted as dots. Also shown are 
AGB stars in Baade's Window from Frogel \& Whitford 
(1987; squares), the Galactic disk from Le Bertre (1992; triangles) and 
Le Bertre (1993; crosses), as well as LPVs in the Galactic Center from Wood et al. 
(1998; stars). Note that the brightest stars near the center of M32 have 
M$_{L'}$ and $K-L'$ that are similar to luminous AGB stars in the Galactic disk.}
\end{figure}

\clearpage

\begin{figure}
\figurenum{5}
\epsscale{0.75}
\plotone{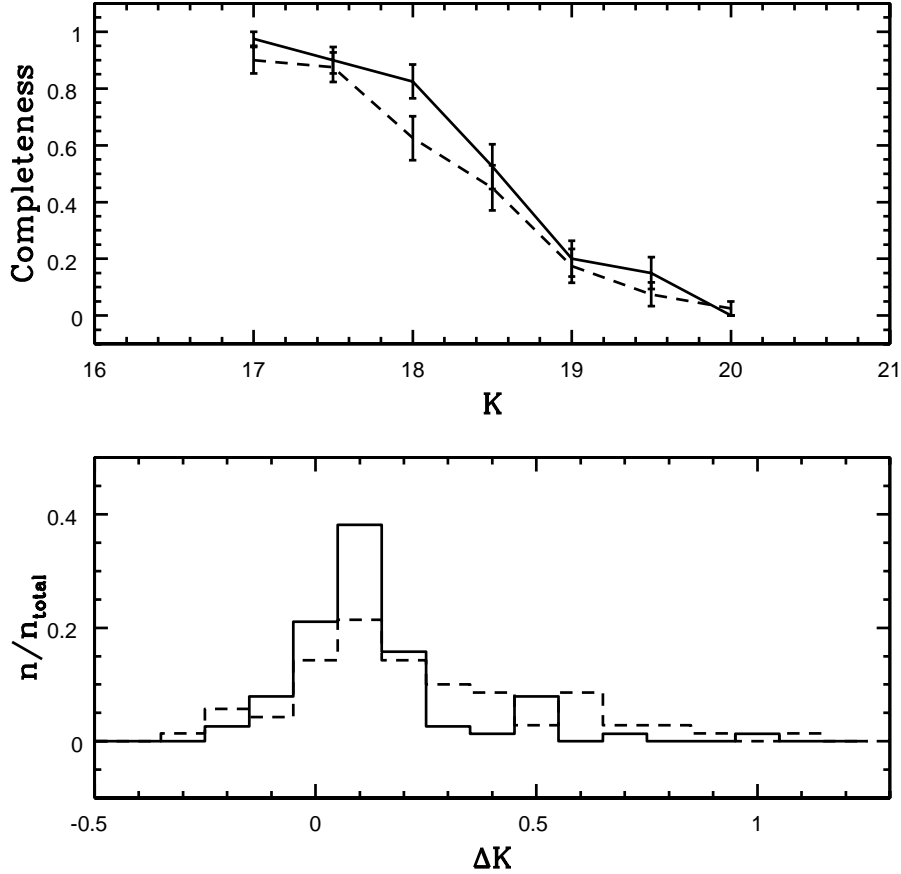}
\caption
{The completeness curves for artificial stars in two annuli in the CFHTIR data are 
compared in the top panel. The dashed line shows the completeness curve for 
objects with projected distances along the major axis between 0.2 and 0.4 kpc, 
while the solid line is for stars between 0.8 and 1.0 kpc. Completeness is 
the ratio of the number of artificial stars recovered to those that were added. 
To be recovered, an artificial star had to be detected in both $H$ and $K'$. 
The tendency for the 0.2 -- 0.4 kpc curve to fall slightly 
below the 0.8 -- 1.0 kpc curve is due to the higher stellar density at smaller 
galactocentric radii. The $\Delta K$ distributions for stars with $K = 18$ 
in the two radial intervals are compared in the lower panel. The distributions have 
been normalized according to the total number of recovered stars in each distance 
interval. The $\Delta K$ distribution for the $0.2 - 0.4$ kpc interval is the broader
of the two, and contains a tail towards large $\Delta K$ values that is 
a consequence of the higher stellar density in this portion of M32.}
\end{figure}

\clearpage

\begin{figure}
\figurenum{6}
\epsscale{0.75}
\plotone{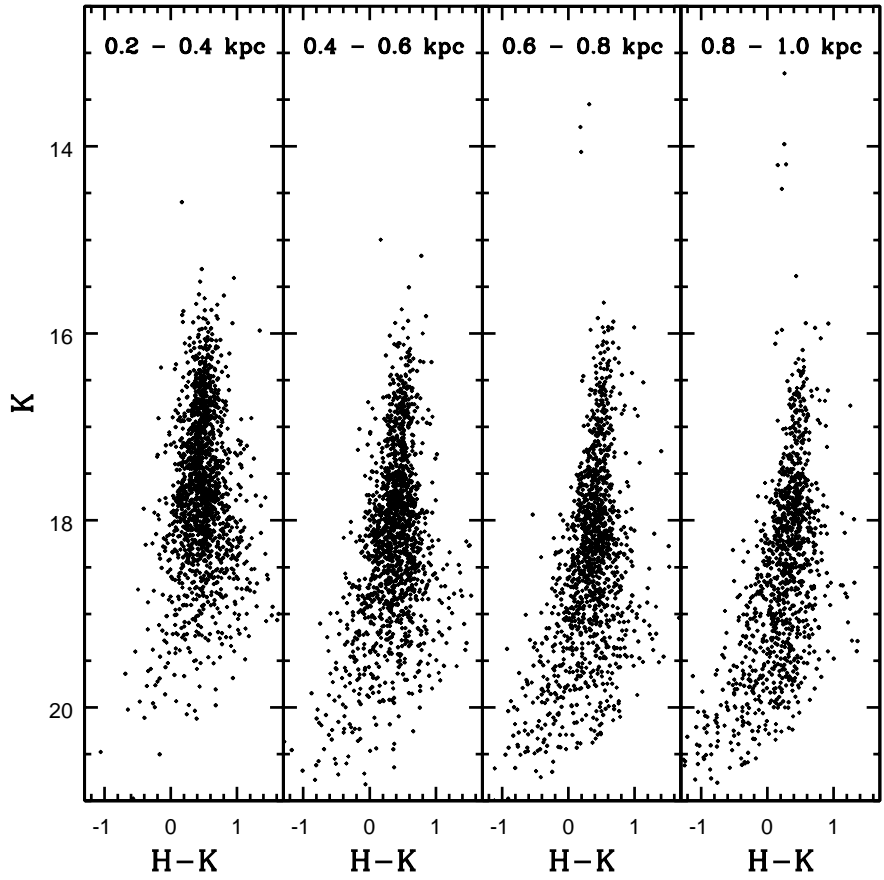}
\caption
{The $(K, H-K)$ CMDs of stars in four radial intervals in the CFHTIR field. 
Distances are along the semi-major axis, assuming a 
distance modulus of 24.5 and an ellipticity of 0.15.} 
\end{figure}

\clearpage

\begin{figure}
\figurenum{7}
\epsscale{0.75}
\plotone{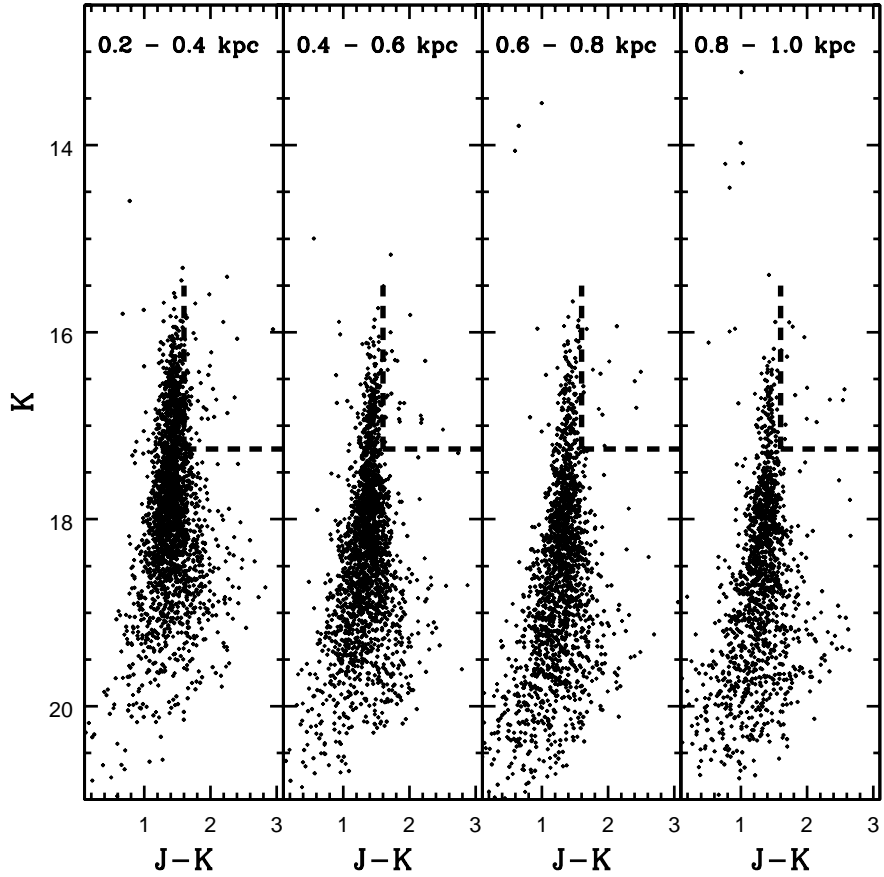}
\caption
{Same as Figure 6, but showing the $(K, J-K)$ CMDs. The dashed lines delineate 
$J-K > 1.6$ and M$_K < -7.25$, which is the region on the CMDs where C stars 
might be expected.} 
\end{figure}

\clearpage

\begin{figure}
\figurenum{8}
\epsscale{0.75}
\plotone{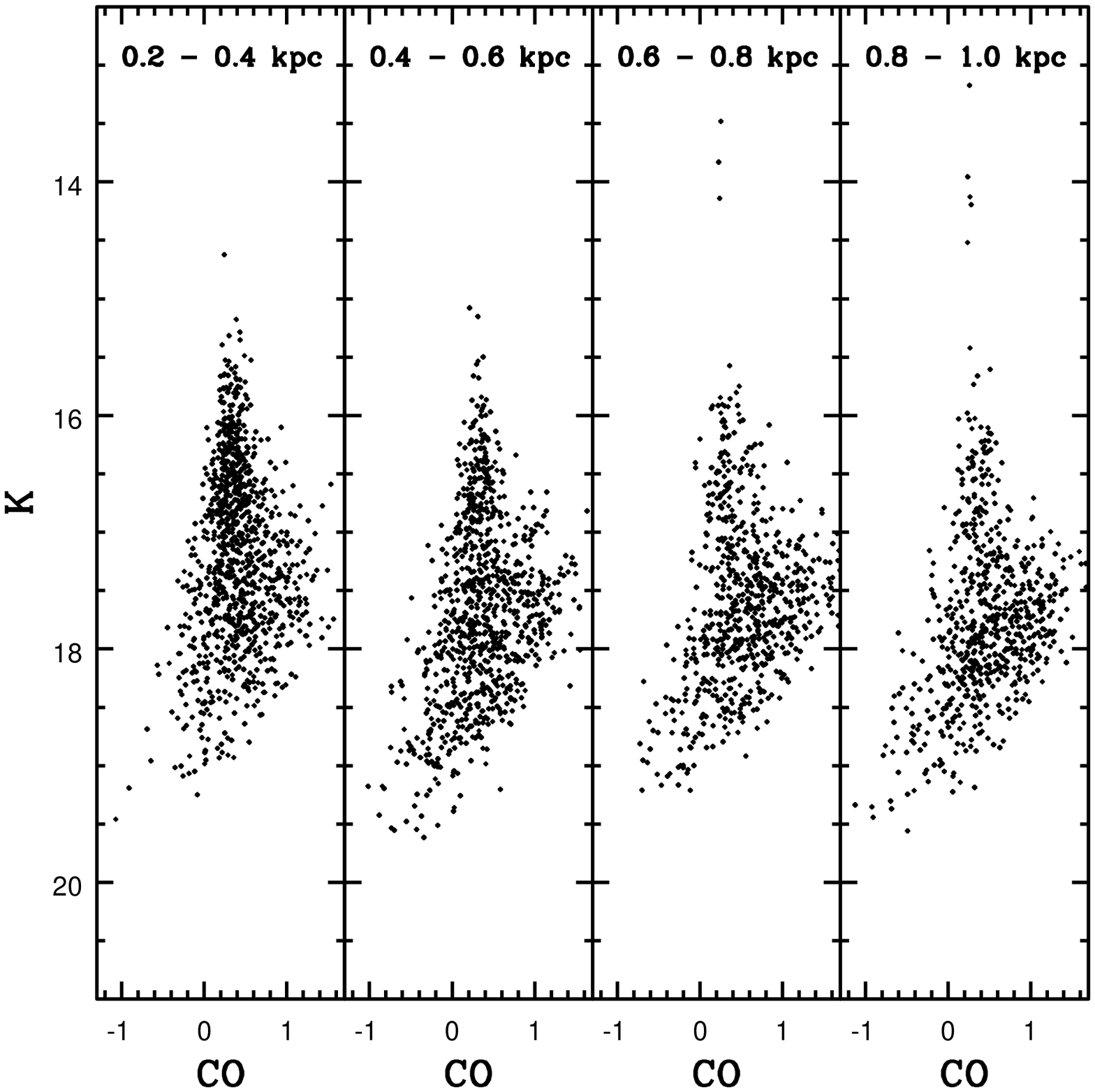}
\caption
{Same as Figure 6, but showing the $(K, CO)$ CMDs.}
\end{figure}

\clearpage

\begin{figure}
\figurenum{9}
\epsscale{0.75}
\plotone{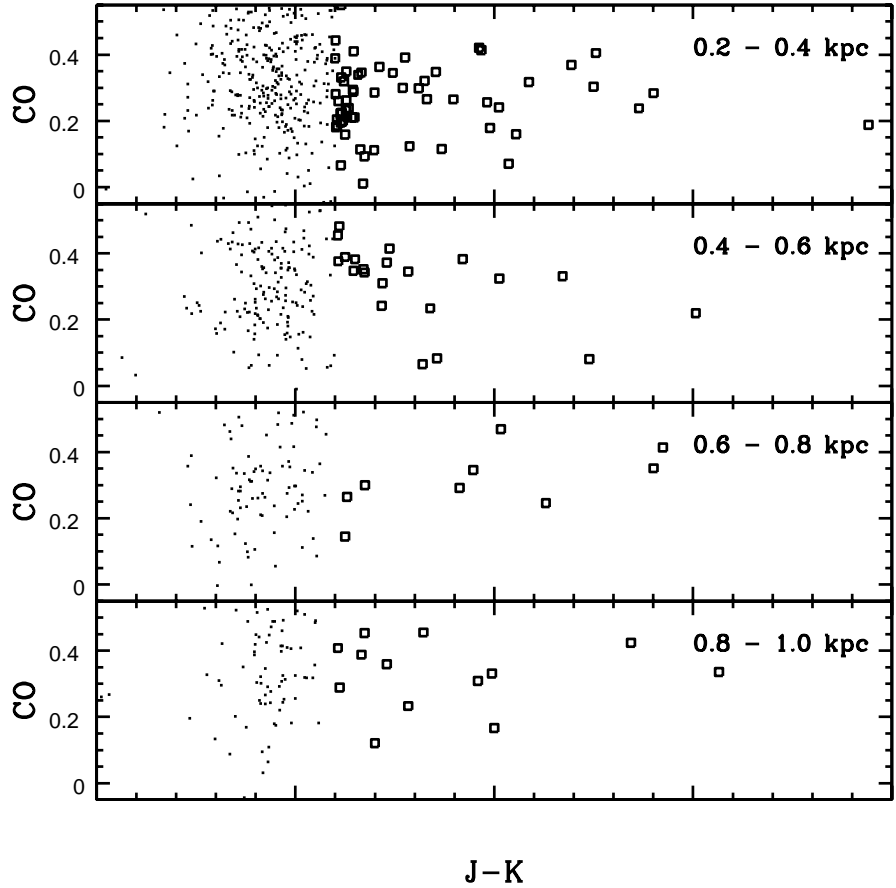}
\caption
{The $(CO, J-K)$ TCD of stars with M$_K < -7.25$ in M32. 
Stars with $J-K > 1.6$ are plotted as open squares, whereas stars with $J-K < 1.6$ are 
shown as dots. Note that the CO index stays roughly constant with $J-K$, and that 
the mean CO index does not change with radius.}
\end{figure}

\clearpage

\begin{figure}
\figurenum{10}
\epsscale{0.75}
\plotone{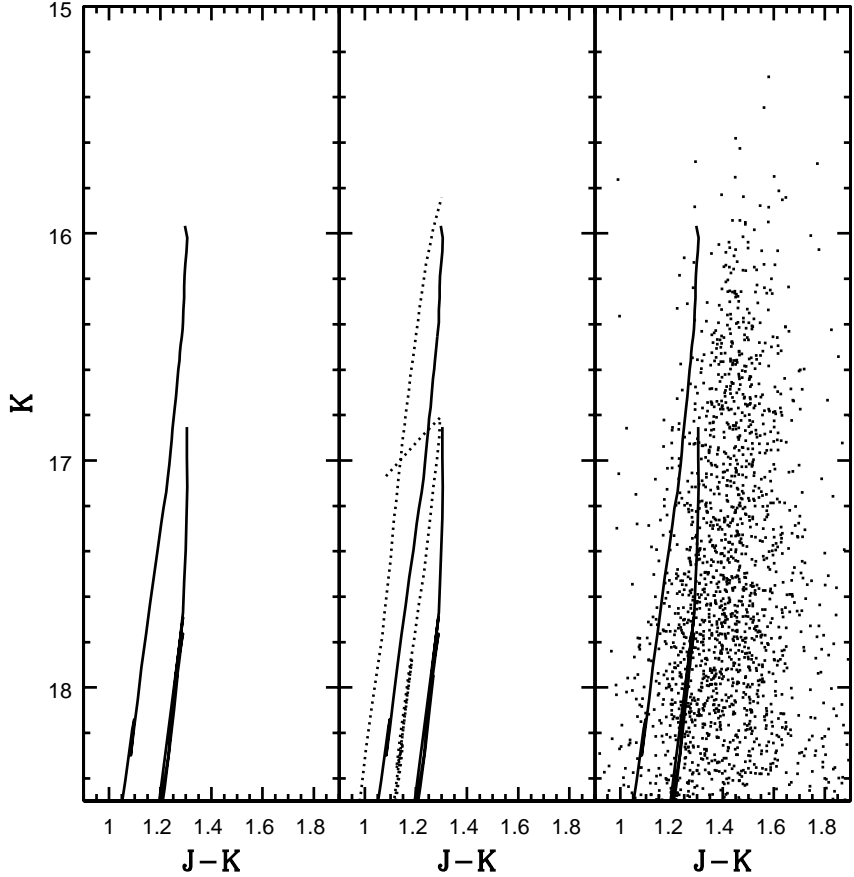}
\caption
{Selected isochrones from Girardi et al. (2002), assuming a distance modulus of 24.5 for 
M32. The sensitivity of the observations to age is investigated in the left hand panel, 
where models with z=0.019 and ages of 1 Gyr and 8 Gyr are compared. The isochrones 
were constructed from the models described in the appendix of Marigo \& Girardi (2001). 
Metallicity sensitivity is investigated in the middle panel, where models with 
z=0.019 (solid lines) and z=0.008 (dashed lines) and ages of 1 and 8 Gyr are compared. 
These comparisons indicate that (1) the AGB-tip brightness in $K$ is 
more sensitive to age than metallicity, and (2) the width of the AGB sequence on the $(K, 
J-K)$ CMD is sensitive to metallicity variations. The z=0.019 models are compared 
with the CFHTIR observations of the 0.2 -- 0.4 kpc 
interval in the right hand panel. The 1 Gyr model 
provides a reasonable match to the observed AGB-tip brightness, although if the brightest 
stars are LPVs near the peak of their light curves then the actual age will be 
older than 1 Gyr.}
\end{figure}

\clearpage

\begin{figure}
\figurenum{11}
\epsscale{0.75}
\plotone{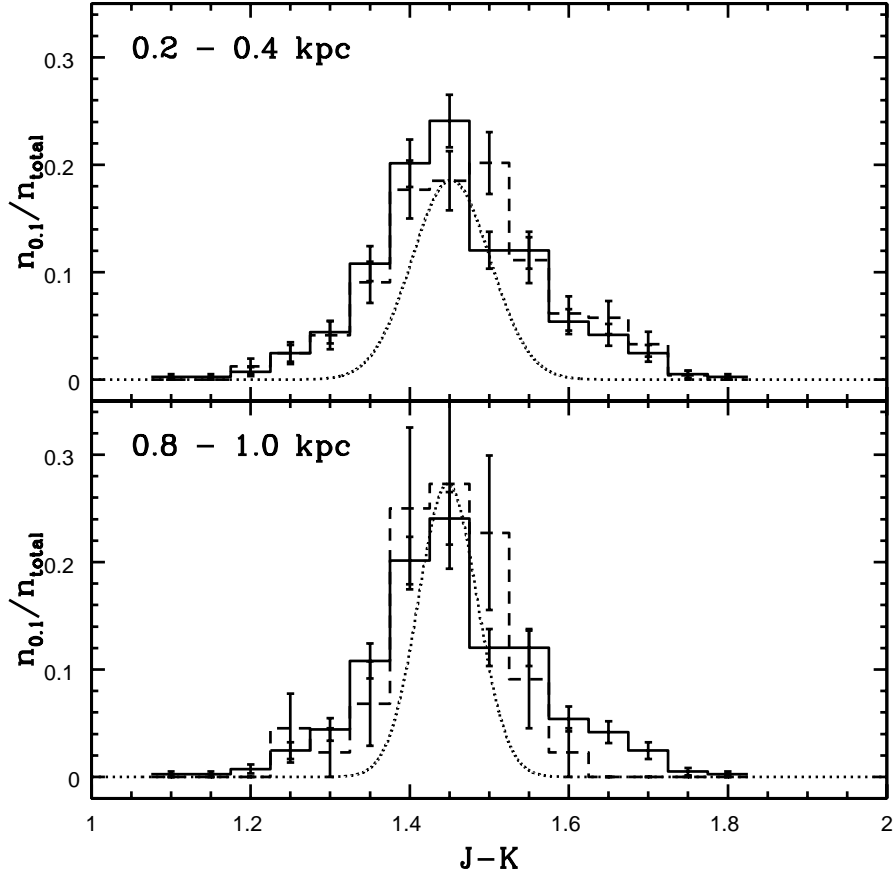}
\caption
{The $J-K$ color distributions for stars with $K$ between 16.5 and 
17.0 in the 0.2 -- 0.4 kpc and 0.8 -- 1.0 kpc intervals are shown 
as dashed lines in the upper and lower panels. The solid line in each 
panel is the color distribution defined by stars 
in this brightness range over the full 0.2 -- 1.0 kpc interval. The distributions 
in each panel have been normalized according to the total number of objects. 
The dotted line in each panel is a gaussian showing the expected dispersion 
due to random photometric errors alone, as determined from artificial star experiments. 
The distributions for both distance intervals are markedly wider 
than expected from photometric errors alone, indicating that factors other than random 
errors in the photometry broaden the distributions.}
\end{figure}

\clearpage

\begin{figure}
\figurenum{12}
\epsscale{0.75}
\plotone{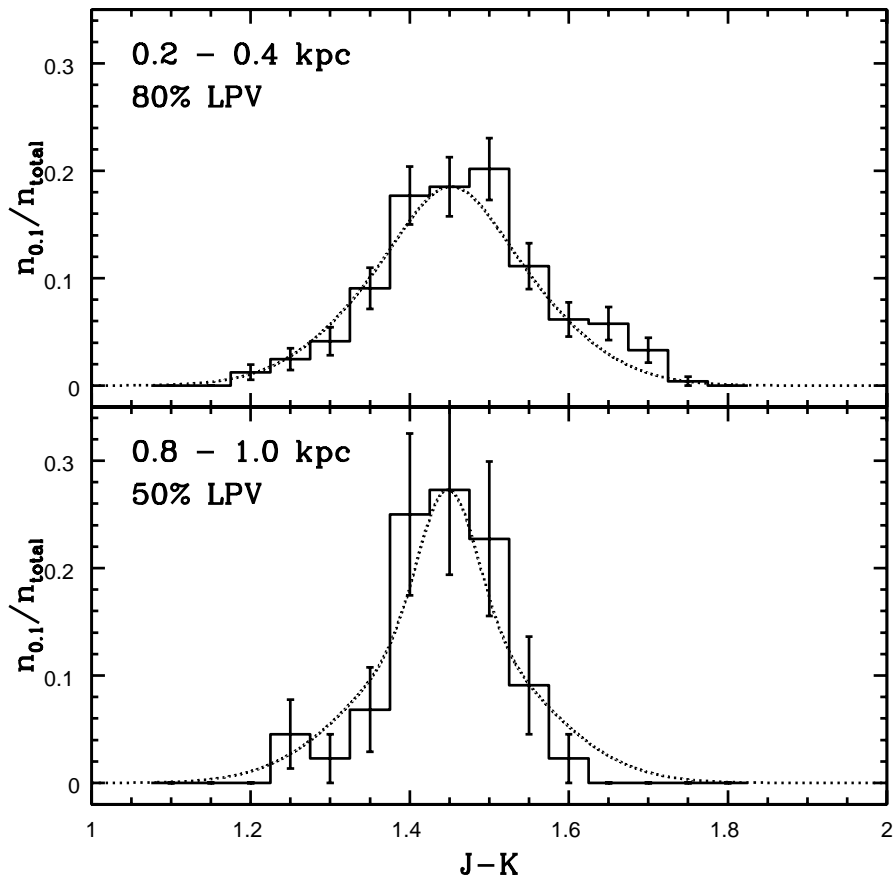}
\caption
{The $J-K$ color distributions from Figure 11 (solid lines) 
are compared with simulated color distributions for a population consisting of LPVs and 
non-variable stars (dotted lines). The model distribution in the top panel assumes 
that LPVs make up 80\% of the AGB content, as was found by Davidge \& 
Rigaut (2004) near the center of M32. Note that while an 80\% LPV fraction provides a 
reasonable match to the 0.2 -- 0.4 kpc distribution, the 
width of the distribution for stars in the 0.8 -- 1.0 kpc interval is 
consistent with $\sim 50\%$ of the stars being LPVs.}
\end{figure}

\clearpage

\begin{figure}
\figurenum{13}
\epsscale{0.75}
\plotone{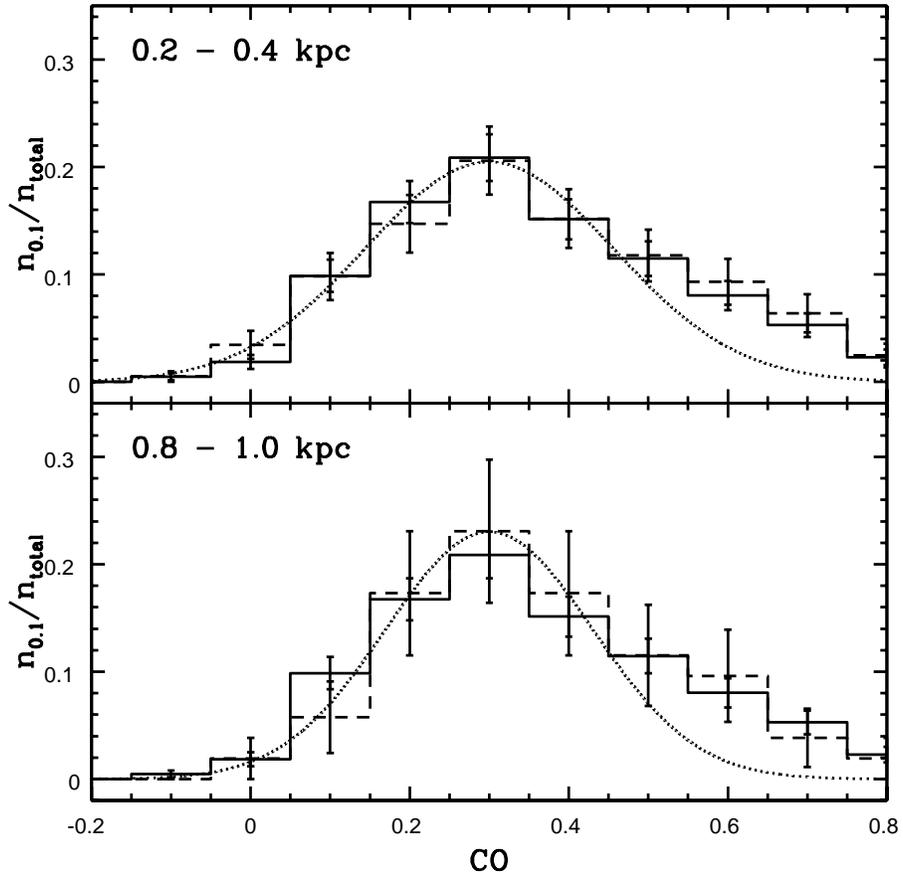}
\caption
{The CO color distributions for the 0.2 -- 0.4 and 0.8 -- 1.0 kpc intervals. The 
identification of the various curves is the same as in Figure 11. The distributions 
have widths that are broadly comparable with photometric errors alone, with a tail of 
CO-strong objects in both distance intervals. Because of the large random photometric 
uncertainties, the CO distributions do not contain significant information about 
differences in stellar content between the two distance intervals.}
\end{figure}

\clearpage

\begin{figure}
\figurenum{14}
\epsscale{0.75}
\plotone{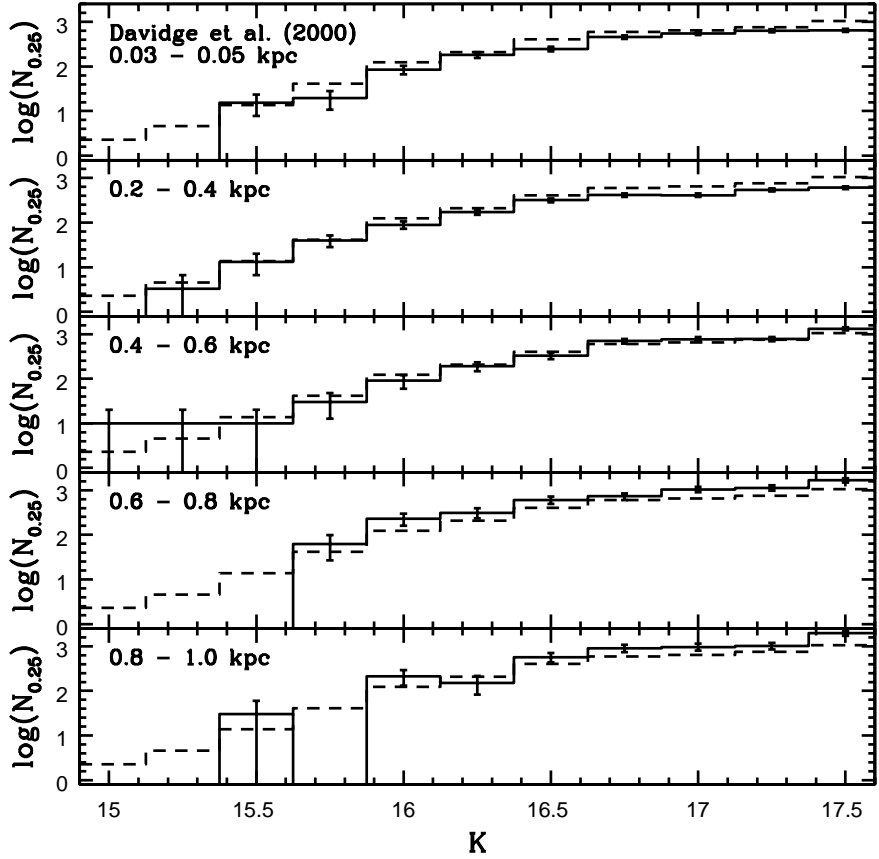}
\caption
{The $K$ LFs of the four distance intervals considered in the CFHTIR data (solid 
lines) compared with the LF of the entire CFHTIR field when $r_{M32} > 0.2$ kpc 
(dashed line). N$_{0.25}$ is the number of stars per 0.25 mag interval in 
$K$, as measured from the $(K, H-K)$ CMD and scaled to 
a total brightness M$_{r} = -15$. The top panel shows the LF of stars with 
projected distances 30 -- 50 parsecs from the center of M32, obtained from 
the data discussed by Davidge et al. (2000). The comparisons in this figure 
indicate that the number of AGB stars scales with integrated $r-$band brightness 
over distances up to 1.0 kpc from the center of M32.}
\end{figure}

\clearpage

\begin{figure}
\figurenum{15}
\epsscale{0.75}
\plotone{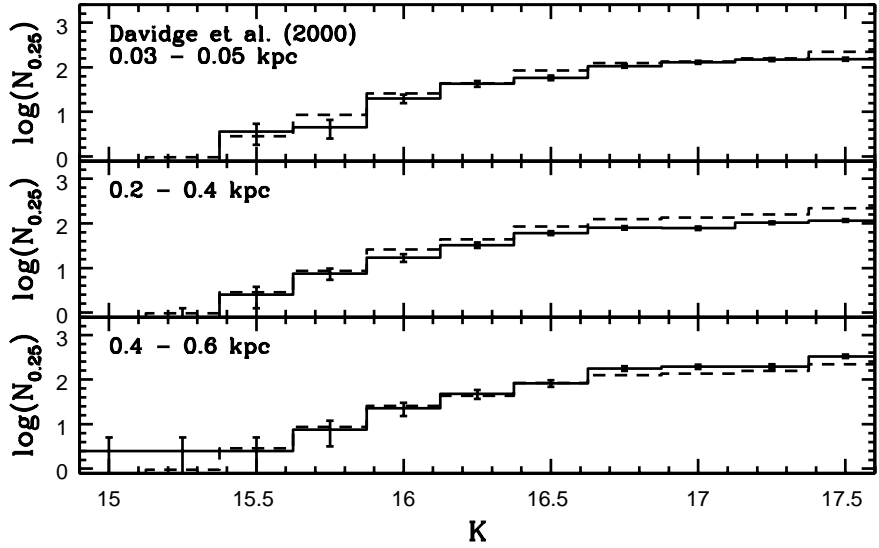}
\caption
{The same as Figure 14, but with the LFs scaled as if each interval sampled a total 
brightness M$_{K} = -16$. Only data for the innermost three annuli are shown, as the 2MASS 
surface brightness profiles of M32 are noisey when $r > 150$ arcsec. The comparisons in 
this figure indicate that the $K$ LF scales with integrated $K$ brightness over 
distances up to at least 0.6 kpc from the center of M32.}
\end{figure}

\clearpage

\begin{figure}
\figurenum{16}
\epsscale{0.75}
\plotone{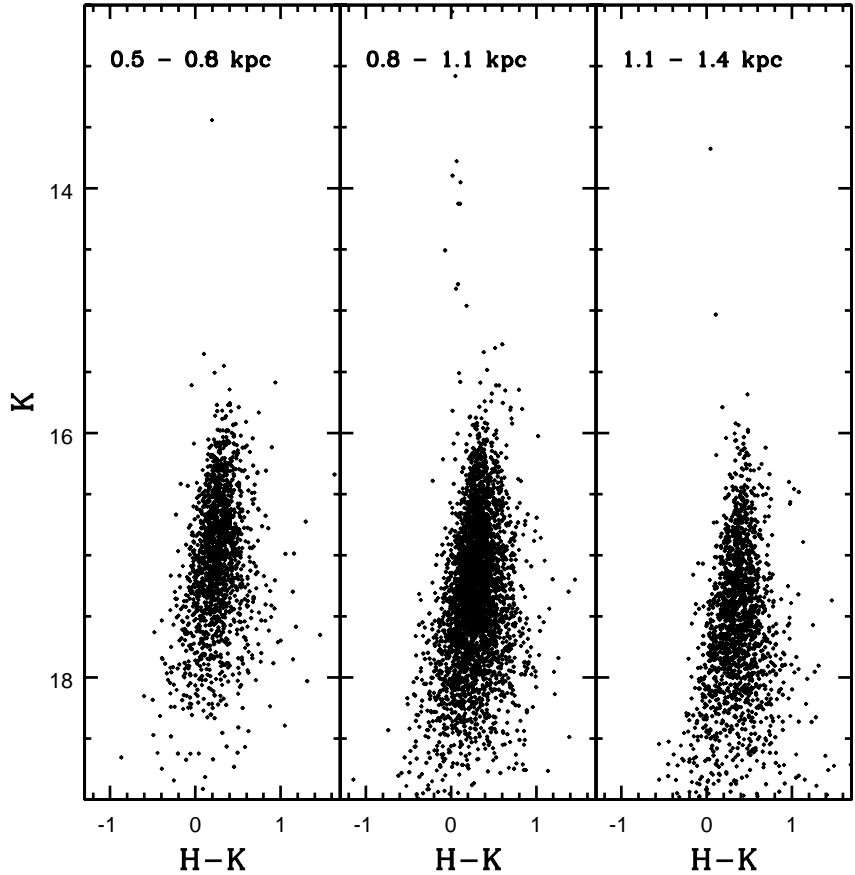}
\caption
{The $(K, H-K)$ CMDs of stars in the M31 CFHTIR field. The distances 
listed are along the semi-minor axis of M31, assuming a 
distance modulus of 24.5 and an ellipticity of 0.26. 
Note the (slight) tendency for the peak $K$ brightness to 
increase towards smaller distances. Artificial star experiments 
suggest that this may be due to blending.}
\end{figure}

\clearpage

\begin{figure}
\figurenum{17}
\epsscale{0.75}
\plotone{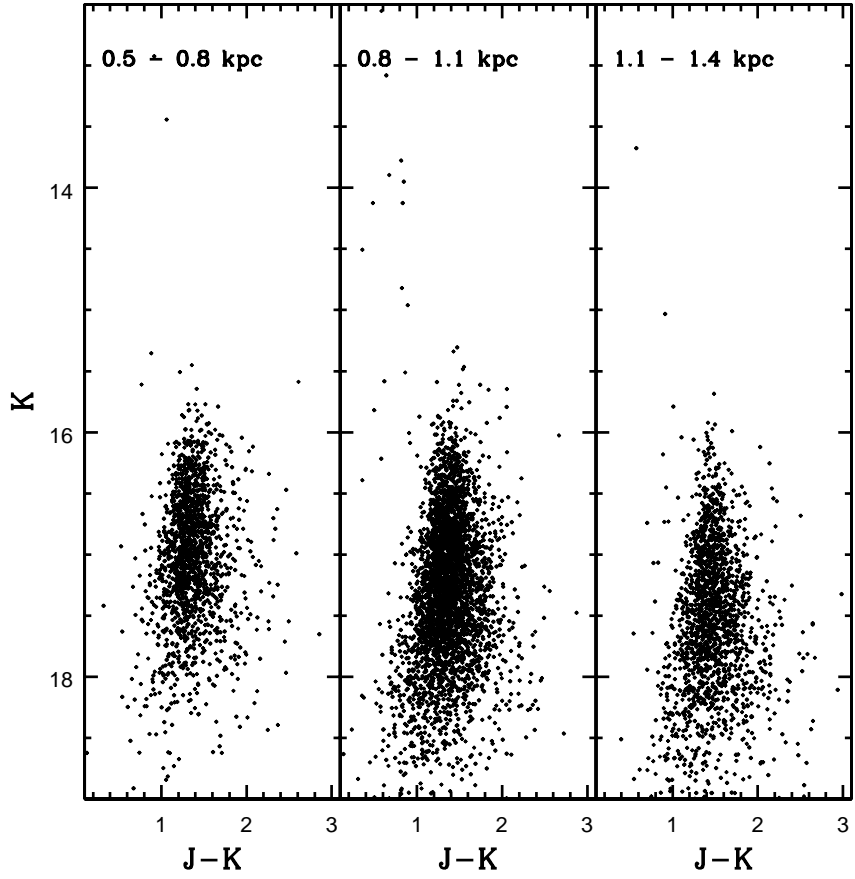}
\caption
{Same as Figure 16, but showing the $(K, J-K)$ CMDs.}
\end{figure}

\clearpage

\begin{figure}
\figurenum{18}
\epsscale{0.75}
\plotone{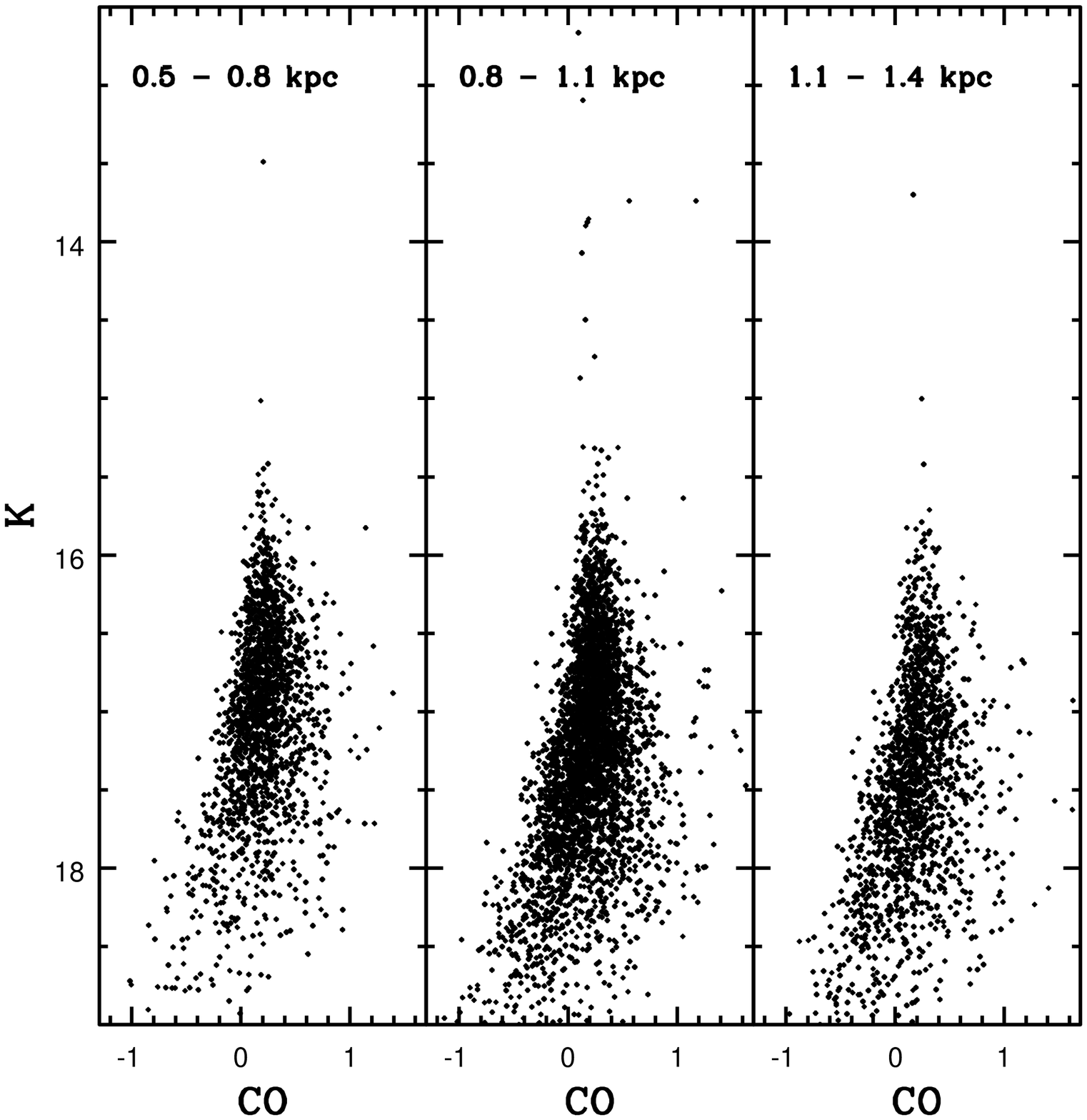}
\caption
{Same as Figure 16, but showing the $(K, CO)$ CMDs.}
\end{figure}

\clearpage

\begin{figure}
\figurenum{19}
\epsscale{0.75}
\plotone{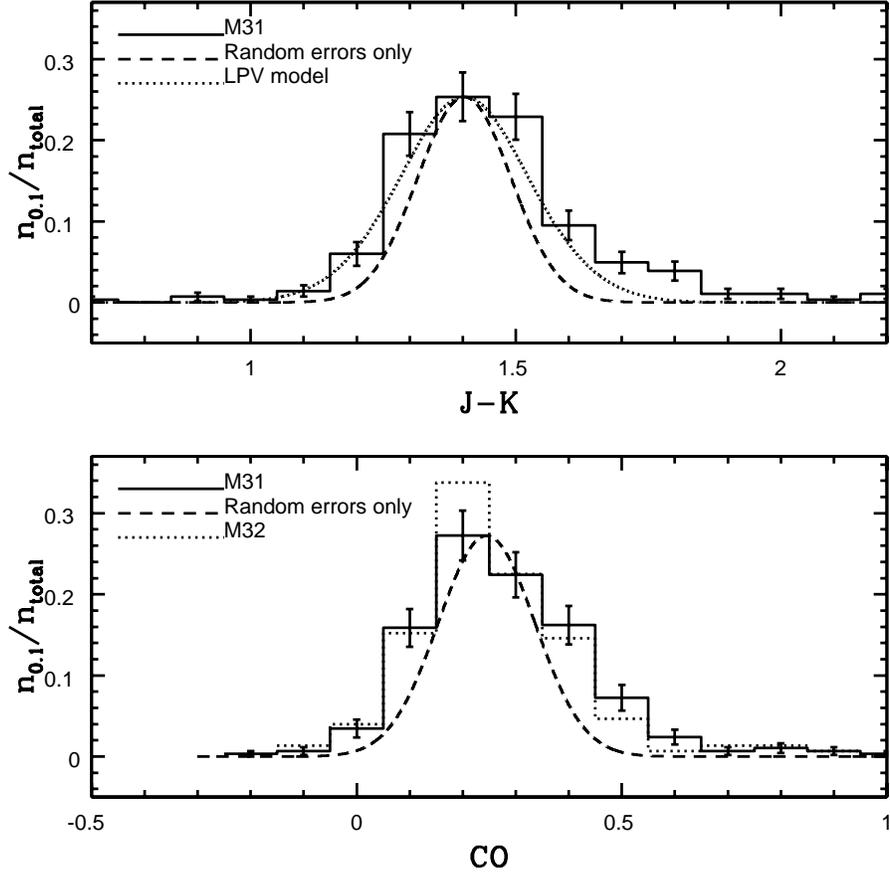}
\caption
{The $J-K$ (top panel) and CO (lower panel) distributions of stars 
with $K$ between 16.5 and 17 located in the 1.1 - 1.4 kpc interval of M31. 
The distributions have been normalized according to the total number of objects in 
each sample. The dashed lines show the gaussian distribution 
predicted from the artificial star experiments, scaled to match the number of objects
in the peak bin. The dotted line in the upper panel shows a model distribution, 
constructed using the procedures described in \S 4, 
in which 80\% of the stars are assumed to be LPVs. The dotted 
line in the lower panel is the CO distribution for stars 
in the 0.2 -- 0.4 kpc interval in M32 with $K$ between 16.5 and 17, 
but shifted to match the peak CO index in the M31 data. 
Note the good agreement between the CO distributions of the two galaxies.}
\end{figure}

\clearpage

\begin{figure}
\figurenum{20}
\epsscale{0.75}
\plotone{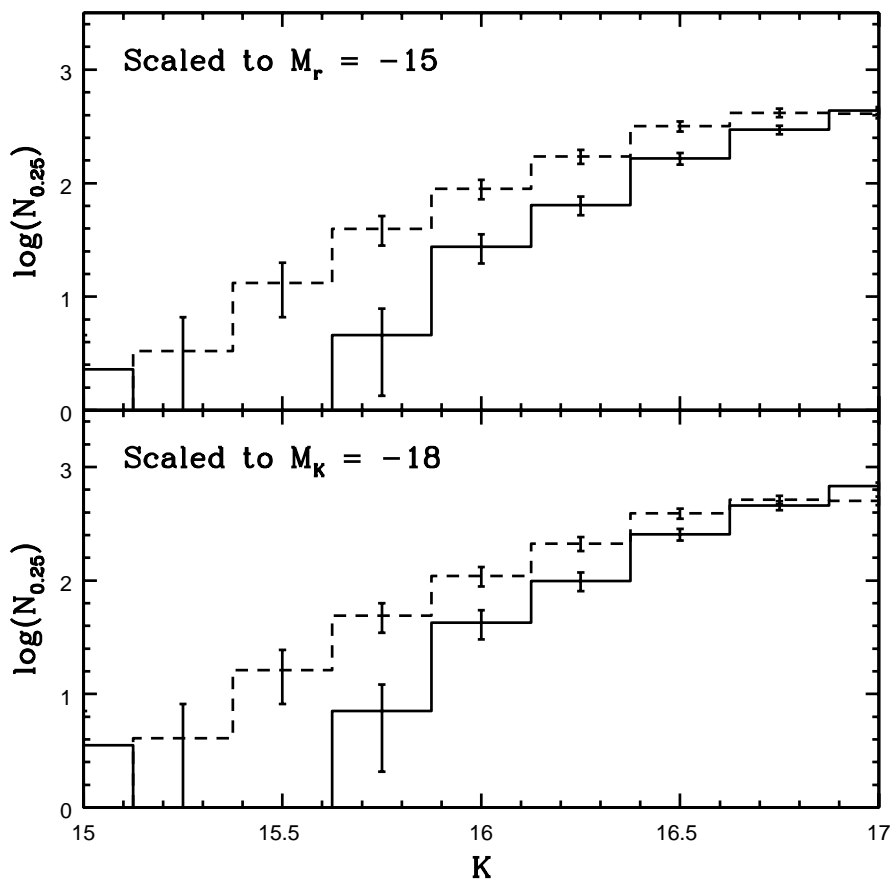}
\caption
{The $K$ LF of stars in the 1.1 - 1.4 kpc interval in the outer bulge of 
M31 (solid line), compared with the LF of stars in the 0.2 -- 0.4 kpc interval in 
M32 (dashed line). N$_{0.25}$ is the number of stars per 0.25 mag 
interval in $K$, scaled as if the integrated brightnesses in the areas studied are 
M$_{r} = -15$ (upper panel) and M$_K = -18$ (lower panel). The error bars 
show the uncertainties due to counting statistics. The comparisons indicate 
that M32 contains more bright AGB stars per unit integrated brightness
than the outer bulge of M31. Note that the differences between the two LFs is 
independent of the wavelength used to normalize the counts.}
\end{figure}

\end{document}